\documentclass[a4paper,11pt]{article}

\usepackage{jheppub} 

\usepackage[T1]{fontenc} 
\usepackage{subfig}
\usepackage{float}
\usepackage{bbold}

\allowdisplaybreaks[1]

\title{Curved spacetime effective field theory (cEFT) -- construction with the heat kernel method}

\author{{\L}ukasz Nakonieczny}

\affiliation{Institute of Theoretical Physics, Faculty of Physics, University of Warsaw \protect \\
ul.~Pasteura 5,~02-093 Warszawa, Poland }

\emailAdd{Lukasz.Nakonieczny@fuw.edu.pl}

\abstract{ 
In the presented paper we tackle the problem of the effective field
theory in curved spacetime (cEFT) construction. To this end, we propose to use the heat kernel method.
After introducing the general formalism based on the well established formulas known from the application
of the heat kernel method to deriving the one-loop effective action in curved spacetime, 
we tested it on selected problems.
The discussed examples were chosen to serve as a check of validity of the derived formulas
by comparing the obtained results to the known flat spacetime calculations. 
On the other hand, they allowed us to obtain new results concerning the influence of the gravity 
induced operators on the effective field theory without unnecessary calculational complications.       
}

\begin{document} 
\maketitle
\flushbottom

\section{Introduction}
\label{sec:intro}

The effective field theory (EFT) turns out to possess an immense usefulness in particle physics. It allows to 
conveniently parametrize the effects coming from the unknown high energy physics and gauge its influence on the 
experimentally measurable observables. Looking at the same problem from a different perspective,
it allows to refine our understanding of the high energy phenomena not yet directly measurable
in experiments by the already obtained indirect data, which are on the theoretical level described by effective operators
For a classification of the flat spacetime operators with a dimension up to six that obey the Standard Model (SM) gauge symmetries
we refer the reader to \cite{BUCHMULLER1986621,Grzadkowski2010}. For some examples of the recent use 
of the EFT in the context of the SM observables calculation see \cite{Dedes2018} and citations therein.            

Having in mind the usefulness of the EFT, it is of considerable importance to have a well tested and 
possibly simple and clear formalism to obtain the effective field theory from a given high energy model. 
Recently, there has been a resurgence of the activity in this area that bore fruits in the form of the 
Covariant Derivative Expansion (CDE) scheme \cite{Henning_Lu_Murayama_2016,Henning_Lu_Murayama_2016_2} and
construction of the Universal Effective Action (UEA) formalism \cite{Drozd_Ellis_Quevillon_You_2016,Ellis2017}.

Meanwhile, the presence of the classical gravitational field described by the curvature of spacetime 
poses new challenges for the quantum field theory. Among them there are questions of the 
influence of gravity on the Standard Model vacuum stability 
\cite{Herranen_Markkanen_Nurmi_Rajantie_2014,Herranen_Markkanen_Nurmi_Rajantie_2015,
Czerwinska_Lalak_Nakonieczny_2015,Rajantie_PRD2018} and the gravity assisted dark matter production 
\cite{Makkanen_Nurmi_2017,Tang_Wu_2017,Artymowski_UW_2018} or bariogenesis 
\cite{majumdar_1995,PhysRevD.60.063513,PhysRevLett.93.201301,LAMBIASE20069,PhysRevD.89.103501,PhysRevD.91.045002,Hamada:2016jnq}. 
To investigate these problems the EFT may be the right tool, yet before this could happen 
it should be reformulated to take into account the spacetime curvature.
This reformulation is the subject of presented article.

To extend the effective field theory into curved spacetime we propose to use 
the heat kernel method \cite{DeWitt_1965,Buchbinder_Odintsov_Shapiro_1992,Avramidi_2000,Parker_Toms_2009}.
The method was already applied with many successes in calculations within the quantum field theory 
in curved spacetime framework. To name a few applications, we list the following problems of: vacuum polarization
\cite{FROLOV1982372,0264-9381-10-3-009,PhysRevD.81.124047,PhysRevD.94.105001},
calculation of the logarithmic divergences and renormalization group (RGE) running of constants for various matter models 
\cite{Buchbinder_Odintsov_Shapiro_1992,Parker_Toms_2009} (and citations therein), 
obtaining renormalization group improved effective action
\cite{Elizalde_Odintsov_1994,Elizalde_Odintsov_1994_2,Elizalde_Kirsten_Odintsov_1994,Elizalde_Odintsov_Romeo_1995,PhysRevD.90.084001,
PhysRevD.91.083529},
one-loop effective action
\cite{Markkanen_Tranberg_2012,Lalak_Nakonieczny_2017,Markkanen2018,Toms2018,PhysRevD.98.025015} 
or the abovementioned question of an influence of gravity on the stability of the Higgs effective potential. 
Additionally, the advocated approach to the cEFT possesses an advantage that the heat kernel method may be viewed as a direct generalization 
of the aforementioned CDE and UEA methods known from flat spacetime to the curved spacetime.

The structure of the article is the following. 
In section \ref{sec:outline} we collected the necessary ingredients that allowed us to use
the heat kernel method to construct the curved spacetime effective field theory (cEFT). 
In section \ref{sec:examples} we used the obtained formalism to work out three examples,
namely the Higgs sector interacting with the heavy scalar singlet, the Yukawa model with the heavy scalar 
and electrically charged fermions. 
In section \ref{sec:summary} we summarized and discussed the obtained results.

\section{Constructing the effective field theory in curved spacetime -- general formulas}
\label{sec:outline}

In this section we will present general formulas relevant for constructing the effective field theory that takes into account effects 
generated by the presence of the heavy matter sector and classical gravitational field. 
In what follows, we will focus on the tree and one-loop contributions form the heavy sector. 
Before we elaborate on the matter part of the action, let us specify the gravity part
\begin{align}
S_g &= \int \sqrt{-g}\ d^4 x \left [  \frac{1}{16 \pi G} \left ( R - 2 \Lambda \right ) + 
\alpha_1 R_{\alpha \beta \mu \nu}R^{\alpha \beta \mu \nu} + \alpha_2 R_{\alpha \beta}R^{\alpha \beta} +\alpha_3 R^2   + 
\alpha_4 \square R\right ].
\end{align}
The first two terms give us the standard Einstein-Hilbert action with the cosmological constant. The terms quadratic in curvatures,
proportional to the $\alpha_i$ coefficients, are introduced in order to obtain the renormalizable gravity sector at the one-loop level 
\cite{Buchbinder_Odintsov_Shapiro_1992,Parker_Toms_2009}.
In what follows, we will be using the $(+++)$ sign convention of \cite{MTW_1973}, this includes mostly plus convention for the metric tensor $(-,+,+,+)$.
Specifying the matter part of the high energy theory $S^{UV}_m$ at this time is not necessary.
Schematically the UV (Ultraviolet) action could be written as
\begin{align}
S^{UV}_{m}(\phi,\Phi) = S^{light}_{m}(\phi) + S^{heavy}_{m}(\Phi) + S^{light,heavy}_{m}(\phi,\Phi),
\end{align}      
where $\phi$ represents light fields (with masses and momenta smaller then some chosen energy scale $\mu_{c}$) and 
$\Phi$ represents heavy fields.
To construct the low energy effective field theory we will use the functional methods. Specifically, we will integrate out heavy fields 
$\Phi$ (see for example \cite{Henning_Lu_Murayama_2016,Henning_Lu_Murayama_2016_2}).  
This gives us the following formal expression for the cEFT containing the one-loop effect coming form the heavy sector
\begin{align}
S^{cEFT}_{m}(\phi) = S^{light}_{m}(\phi) + S^{light,heavy}_{m}(\phi,\Phi)_{| \Phi = \Phi_{cl}(\phi)} + 
\frac{i \hbar}{2} c_s \ln sdet \left ( \mu^{-2} D^2_{i j} \right )_{| \Phi = \Phi_{cl}(\phi)},
\end{align} 
where $\Phi_{cl}(\phi)$ is the classical (tree-level) solution to the heavy fields equations of motion 
$\frac{\delta S^{UV}_{m}}{\delta \Phi} =0$, $\frac{\delta}{\delta \Phi}$ represents the functional derivative of $S^{UV}_{m}$
and $c_s$ is the usual spin dependent coefficient, for example $c_s = +1$ for a real scalar and $c_s = -2$ for the Dirac fermions. 
The symbol $sdet$ represents a functional superdeterminant of the operator $D^2$ and $\mu^2$ is some arbitrary energy scale introduced 
to make the argument of the determinant dimensionless. The operator $D^2$ is constructed from the UV matter action as 
\begin{align}
\label{D2_min}
D^2_{ij} \equiv \frac{\delta^2 S^{UV}_{m}(\phi, \Phi)}{\delta \Phi_i \delta \Phi_j}.
\end{align}
In the above formula we restrained ourselves to taking into account only an effect of the heavy particle loops.

To give a meaning to the formal expression $\ln sdet \left ( \mu^{-2} D^2 \right )$ we will use the heat kernel method
(within the Schwinger-DeWitt approximation) \cite{Buchbinder_Odintsov_Shapiro_1992,Avramidi_2000,Parker_Toms_2009}.
From now on we will assume that the operator defined by (\ref{D2_min}) is of the form
\begin{align}
\label{operatorD2}
D^2 = \square + 2 h^{\mu}(\phi,\Phi_{cl}(\phi))d_{\mu} + \Pi(\phi,\Phi_{cl}(\phi)) - m_{\Phi}^2,
\end{align}
where $\square \equiv d_{\mu} d^{\mu}$ is the d'Alembert operator, $d_{\mu} \equiv \nabla_{\mu} + i e_s A_{\mu}$
is a covariant derivative containing a gauge part $A_{\mu}$ with $e_s$ being the charge of the field it acts upon 
and a gravity part encapsulated in an ordinary covariant derivative defined in curved spacetime $\nabla_{\mu}$, 
moreover, $m_{\Phi}^2$ is a positive constant that could be equated to the heavy particle mass.
$\Pi(\phi,\Phi_{cl}(\phi))$ is the part that does not contain any open (acting on non-background fields) 
covariant derivatives.\footnote{It is worthy to point out that splitting between parts of $D^2$ that does not contain open derivatives
among $\Pi(\phi,\Phi_{cl}(\phi))$ and $m_{\Phi}^2$ terms is somewhat arbitrary. For example, if the $\Phi$ field 
would be in the symmetry broken phase it would be advantageous to promote the $m_{\Phi}^2$ to be the field dependent mass 
$m_{\Phi}^2 \rightarrow m_{\Phi}^2 + f(\Phi)$, where the precise form of $f$ depends on the form of the Lagrangian.
The requirement is that we should have $m_{\Phi}^2 > 0$ for (\ref{Gamma_nr}) to be valid.}
As a side note, let us point that if the $UV$ action is renormalizable in the flat spacetime sense
the heavy scalar fields naturally lead to the above form of the operator while for Dirac fields 
we may achieve this for example by suitable field redefinition in the path integral \cite{Buchbinder_Odintsov_Shapiro_1992}.  
On the other hand, for the gauge fields we may bring the resulting operator to this form by suitable choice of the gauge, yet
in this case we should remember about possible gauge dependence of the obtained results.
Let us also point out that a generalization of the Schwinger-DeWitt technique to the case of operators of more general 
form also exists, see for example \cite{Barvinsky_Vilkovisky_1987,Barvinsky_Vilkovisky_1990}.    
Working within the dimensional regularization framework we arrive at the following formula:
\begin{align}
\label{Gamma_nr}
\Gamma^{(1)}_{\Phi \Phi} &\equiv \frac{i \hbar}{2} c_s \ln sdet \left ( \mu^{-2} D^2 \right ) = 
 \nonumber \\
& = c_s   \int \sqrt{-g}\ d^4 x \frac{\hbar}{64 \pi^2} Tr \bigg \{
a_{0} m_{\Phi}^4 \bigg [ \frac{2}{\bar{\varepsilon}} 
- \ln \bigg ( \frac{m_{\Phi}^2}{\mu^2} \bigg )  + \frac{3}{2} \bigg ]
- 2 a_{1} m_{\Phi}^2 \bigg [ 
\frac{2}{\bar{\varepsilon}} + 1 - \ln \left ( \frac{m_{\Phi}^2}{\mu^2} \right )
\bigg ] + \nonumber \\
&+ 2 a_{2} \bigg [ \frac{2}{\bar{\varepsilon}} - \ln \bigg ( \frac{m_{\Phi}^2}{\mu^2} \bigg )  \bigg ]
+ 2 \sum_{k \geq 3} \frac{a_{k}}{ k(k-1)(k-2) m_{\Phi}^{2(k-2)} }
\bigg \},
\end{align}
where $\frac{2}{\bar{\varepsilon}} = \frac{2}{\varepsilon} - \gamma + \ln(4\pi)$, $\gamma$ is the Euler constant,
the number of spacetime dimensions is $n = 4 - \varepsilon$
and $Tr$ stands for the matrix $tr$ and a sum over all discrete indices (group or Lorentz ones). 
We would like to note here that the formula (\ref{Gamma_nr}) represents a local approximation to the one loop part to the effective action,
therefore it does not contain information about non-local phenomena like for example particle production
\cite{PhysRev.82.664,PhysRevD.20.1772,PhysRevD.21.2756,PhysRevD.60.104045}.
Nevertheless, as far as the effective field theory is concerned it is particularly well suited for expressing
effects of the heavy field in terms of the higher dimensional operators.    
On the other hand, for some example of the use of the non-local form of the heat kernel to the 
construction of the effective action we refer the reader to \cite{Codello2016} and citations therein.

The quantities $a_{k}$ present in (\ref{Gamma_nr}) are the Hadamard-DeWitt (HDW) coefficients \cite{DeWitt_1965}. 
To study the influence of operators up to dimension six in the effective field theory we need
coefficients $a_0$ through $a_{3}$ and some part of $a_4$ containing gravity induced operators of suitable dimension.
Appropriate coefficients are given by \cite{Avramidi_2000} (we will closely follow the notation presented there):
\begin{align}
\label{coef_a0}
a_0 &= 1, \\
a_1 &\equiv P  = Q  + \frac{1}{6}R , \\
a_2 &= P^2 + \frac{1}{3} Z_{(2)}, \\
\label{coef_a3}
a_3 &= P^3 + \frac{1}{2} \left \{ P, Z_{(2)} \right \} + \frac{1}{2} B^{\mu} Z_{\mu} + \frac{1}{10} Z_{(4)},
\end{align}
where $\left \{ ~~,~~ \right \}$ stands for the anticomutator and 
\begin{align}
\label{auxiliary_q_0}
Q &= \Pi - d_{\mu}h^{\mu} - h_{\mu}h^{\mu}, \\
Z_{\mu} &= d_{\mu} P - \frac{1}{3}J_{\mu}, \\
B_{\mu} &= d_{\mu} P + \frac{1}{3}J_{\mu}, \\
J_{\mu} &= d_{\alpha} W^{\alpha}_{\mu}, \\
W_{\alpha \beta} &= [d_{\alpha}, d_{\beta}] - 2 d_{[ \alpha} h_{\beta ]} - 2 h_{[ \alpha} h_{\beta ]}, \\
Z_{(2)} &= \square \left ( Q + \frac{1}{5} R \right ) +
 \frac{1}{30} \left ( R_{\alpha \beta \gamma \delta } R^{\alpha \beta \gamma \delta } - R_{\mu \nu} R^{\mu \nu} \right )
 + \frac{1}{2} W_{\alpha \beta} W^{\alpha \beta}, \\
 Z_{(4)} &= Q_{(4)} + 2 \left \{  W^{\mu \nu} , d_{\mu} J_{\nu} \right \} + \frac{8}{9} J_{\mu} J^{\mu}
 + \frac{4}{3} d_{\mu} W_{\alpha \beta} d^{\mu} W^{\alpha \beta}  + \nonumber \\
 &+ 6 W_{\mu \nu} W^{\nu }_{~~ \gamma} W^{\gamma \mu}
 + \frac{10}{3} R^{\alpha \beta} W^{\mu}_{~~\alpha} W_{\mu \beta} - R^{\mu \nu \alpha \beta}W_{\mu \nu} W_{\alpha \beta}
 + \nonumber \\
 &+ \bigg \{ \frac{3}{14} \square^2 R + \frac{1}{7} R^{\mu \nu} d_{\mu} d_{\nu} R - \frac{2}{21} R^{\mu \nu} \square R_{\mu \nu}
+ \frac{4}{7} R^{\alpha~~\beta}_{~~\mu~~\nu} d_{\alpha} d_{\beta} R^{\mu \nu}
 + \nonumber \\
 & + \frac{4}{63} d_{\mu} R d^{\mu}R  - \frac{1}{42} d_{\mu} R_{\alpha \beta} d^{\mu} R^{\alpha \beta}
 - \frac{1}{21} d_{\mu} R_{\alpha \beta}d^{\alpha} R^{\beta \mu}
+ \frac{3}{28} d_{\mu} R_{\alpha \beta \gamma \delta} d^{\mu} R^{\alpha \beta \gamma \delta}
 + \nonumber \\
 & + \frac{2}{189} R^{\alpha}_{~~\beta} R^{\beta}_{~~\gamma}R^{\gamma}_{~~\alpha}
 - \frac{2}{63} R_{\alpha \beta} R^{\mu \nu} R^{\alpha~~\beta}_{~~\mu~~\nu}
+ \frac{2}{9} R_{\alpha \beta}R^{\alpha}_{~~\mu \nu \lambda } R^{\beta \mu \nu \lambda}
 + \nonumber \\
 & - \frac{16}{189} R_{\alpha \beta}^{~~~~\mu \nu} R_{\mu \nu }^{~~~~\sigma \rho}R_{\sigma \rho}^{~~~~\alpha \beta}
 - \frac{88}{189} R^{\alpha~~\beta}_{~~\mu~~\nu}R^{\mu~~\nu}_{~~\sigma~~\rho}R^{\sigma~~\rho}_{~~\alpha~~\beta}
 \bigg \}, \\
 \label{auxiliary_q_3}
 Q_{(4)} &= \square^2 Q - \frac{1}{2} \left [ W^{\mu \nu}, \left [ W_{\mu \nu} ,Q \right ] \right ]
 - \frac{2}{3} \left [ J^{\mu}, d_{\mu} Q \right ] + \frac{2}{3} R^{\mu \nu} d_{\mu} d_{\nu}Q
 + \frac{1}{3} d_{\mu}R d^{\mu}Q.
\end{align}
The expression for $a_4$ is too unwieldy to be presented here, so we skip it for now, its full form can be found in \cite{Avramidi_2000}
and we added its part containing operators of the dimension six or lower and terms up to the order $O(\mathcal{R}^2)$ in 
Appendix \ref{appA}.
Returning to (\ref{Gamma_nr}) and adopting the $\overline{MS}$ renormalization scheme with the
choice of the running energy scale $\mu^2 = m_{\Phi}^2$ we may obtain
\begin{align}
\label{Gamma_ren}
\Gamma^{(1)}_{\Phi \Phi} = c_s  \int \sqrt{-g} d^4 x 
\frac{\hbar}{64 \pi^2} Tr \left \{ 
\frac{1}{3} \frac{a_3}{m_{\Phi}^2} + \frac{1}{12} \frac{a_{4}}{m_{\Phi}^4}
\right \},
\end{align}
where we took into account the fact that terms present in the $a_1$ coefficient are of the same type as these
in the tree level action, therefore they lead only to the renormalization of the tree level couplings.
As has already been pointed above, we will need only some parts of the $a_{4}$ coefficient.



\section{Effective field theory in curved spacetime -- examples}
\label{sec:examples}

In this section we will present the results concerning an application of the selected scheme of creating the effective field theory
in curved spacetime to some examples. In the case when their flat spacetime counterparts are known they will
serve as a check of validity for our formulas. On the other hand, they will also allow us to 
present some new results illustrating how the presence of the gravitational field modifies the effective field theory.  

\subsection{Singlet scalar interacting with the Higgs sector}
\label{subsec:singlet}

We begin by writing a concrete form of the matter part of the UV theory
\begin{align}
\label{actionUV_XH}
S^{UV}_{m} &= \int \sqrt{-g} d^4 x \bigg(
- \frac{1}{2} d_{\mu} H^{\dagger} d^{\mu} H - \frac{1}{2} m_{H}^2 |H|^2 - \frac{\lambda_{H}}{4!} |H|^4 - \xi_{H} R |H|^2 
+ \nonumber \\
&- \frac{1}{2} d_{\mu} X d^{\mu} X - \frac{1}{2} m_{X}^2 X^2 - \xi_X R X^2 - \frac{1}{3!} m_{3X} X^3
 - \frac{\lambda_{X}}{4!} X^4 
+ \nonumber \\
&- m_{X \xi} X R - m_3 X 
- m_{HX} X |H|^2 - \frac{1}{2} \lambda_{HX} X^2 |H|^2
\bigg),
\end{align}
where $H$ is the Standard Model Higgs doublet, $d_{\mu}$ is a covariant derivative containing gauge fields parts.
For the case where $X$ represents the heavy scalar singlet with mass $m_{X}^2 > 0$ 
(we assume the following mass hierarchy: $m_{X}^2 >>  | m_{H}^2|$) 
$d_{\mu}$ reduces to the standard covariant derivative in curved spacetime  $\nabla_{\mu}$.
Since for now we want to keep the coupling among $X$ and the Higgs doublet described by the term $m_{HX} X |H|^2$
we also need the two remaining terms linear in $X$ if we want our model to be renormalizable. 

In the first step we will solve the classical equation of motion for the $X$ field, actually in what follows we will need only
the solution to the linearized equation of motion \cite{Henning_Lu_Murayama_2016}
\begin{align}
\left ( \square - m_{X}^2 - 2 \xi_X R - \lambda_{HX} |H|^2 \right ) X = m_{HX} |H|^2 + m_{X\xi}R + m_3.
\end{align}  
The formal solution to this is given by
\begin{align}
X_{cl}(|H|^2) = \frac{1}{\square - m_{X}^2 - (\lambda_{XH} |H|^2 + 2 \xi_X R)} 
\left (  m_{HX} |H|^2 + m_{X\xi}R + m_3 \right ).
\end{align}
Expanding it in the large mass limit we get
\begin{align}
X_{cl}(|H|^2) = &- \frac{1}{m_{X^2}} \bigg ( 
1 + \frac{\square - \lambda_{HX}|H|^2 - \xi_X R}{m_{X^2}} + \nonumber \\
&+ \frac{\square - \lambda_{HX}|H|^2 - \xi_X R}{m_{X^2}}  \frac{\square - \lambda_{HX}|H|^2 - \xi_X R}{m_{X^2}} 
+ ... \bigg ) \times \nonumber \\
& \left (  m_{HX} |H|^2 + m_{X\xi}R + m_3 \right ),
\end{align}
where $ + ...$ stands for terms that would produce effective operators of a dimension greater then six.
Keeping only operators of dimension six or less and containing at most terms of the second order in curvatures or
fourth derivatives of the metric we may write 
\begin{align}
\label{Xcl_cur}
X_{cl}(|H|^2) = -\frac{1}{m_{X}^2} \bigg \{
& \bigg [
m_{HX} R + m_3 + \frac{m_{X\xi}}{m_{X}^2} \square R + 2 \frac{m_{X \xi}}{m_{X}^2} \xi_X R^2 + \nonumber \\
 &+ 2\frac{m_3}{m_{X}^2} \xi_X R + 4 \frac{\xi_{X}^2 m_3}{m_{X}^4} R^2 + 2 \frac{m_3 m_{X \xi}}{m_{X}^4} \xi_X \square R
\bigg ] + \nonumber \\
+ |H|^2 & \bigg [ m_{HX} + 2\frac{m_{HX}}{m_{X}^2} \xi_X R + 2 \frac{m_{X\xi}}{m_{X}^4} \xi_X R^2 + 
\frac{\lambda_{HX} m_3}{m_{X}^2} + \nonumber \\
&+ 4 \frac{m_{HX}}{m_{X}^4} \xi_X^2 R^2 + 2 \frac{m_{HX}}{m_{X}^4} \xi_X \left \{ \square , R \right \} +\nonumber \\
&+ 4 \frac{\lambda_{HX} \xi_X}{m_{X}^4} \left ( m_{X \xi} R + m_3 \right ) R 
+ \frac{\lambda_{HX} m_{X \xi}}{m_{X}^4} \square R
\bigg ] + \nonumber \\
+ |H|^4 &  \bigg [
\frac{\lambda_{HX} m_{HX}}{m_{X}^2} + 4 \frac{m_{HX}}{m_{X}^4} \xi_X \lambda_{HX} R + 
\frac{\lambda_{HX}^2}{m_{X}^4} \left ( m_{X \xi} R + m_3 \right )
\bigg ] + \nonumber \\
+ |H|^6 & \bigg [  \lambda_{HX} \frac{m_{HX}}{m_{X}^4} \bigg ] + \nonumber \\
+ \square |H|^2 & \bigg [
\frac{m_{HX}}{m_{X}^2} + 2 \frac{m_{HX}}{m_{X}^4} \xi_X R + \frac{m_3}{m_{X}^4} \lambda_{HX} 
\bigg ] + \nonumber \\
+ \square \left ( |H|^2 R \right ) & \bigg [ \frac{\lambda_{HX} m_{X \xi}}{m_{X}^4} + 2 \frac{m_{HX}}{m_{X}^4} \xi_X
\bigg ] + \nonumber \\
+ \square^2 |H|^2 & \bigg [ \frac{m_{HX}}{m_{X}^4} \bigg ] + \nonumber \\
+ \square |H|^4 & \bigg [  \lambda_{HX} \frac{m_{HX}}{m_{X}^4} \bigg ] + \nonumber \\
+ |H|^2 \square |H|^2 & \bigg [ \lambda_{HX} \frac{m_{HX}}{m_{X}^4} \bigg ]
\bigg \}
\end{align}
To make our notation more concise we may rewrite the above formula as
\begin{align}
X_{cl}(|H|^2) &= a_{H0} |H|^0 + a_{H2} |H|^2 + a_{H4} |H|^4 + a_{H6} |H|^6  + a_{2d H2} \square |H|^2  + \nonumber \\
&+ a_{2d R H2} \square \left ( R |H|^2 \right ) + a_{4d H2} \square^2 |H|^2 + a_{2d H4} \square |H|^4
+ a_{H2 2d H2} |H|^2 \square |H|^2.
\end{align}
In the next step we may use the classical equation of motion for the $X$ field to reduce the tree-level $X$ dependent part of the action
to the form
\begin{align}
S^{Tree}_{X~on~shell} = \int \sqrt{-g} d^4 x \bigg(
\frac{1}{12} m_{3X}X^3 + \frac{1}{24} \lambda_X X^4 -\frac{1}{2} m_{X\xi} X R - \frac{1}{2} m_3 X 
- \frac{1}{2} m_{HX} X |H|^2
\bigg).
\end{align}
From now on we will require that there are no sources for the $X$ field other than the one 
coming form interactions with  other fields. This implies $m_{3} = m_{X \xi} = 0$.
In this case the only terms that will contribute to the effective action for the light field will be
\begin{align}
S^{Tree}_{X~on~shell} = \int \sqrt{-g} d^4 x \bigg(\frac{1}{12} m_{3X} X^3 + 
\frac{1}{24} \lambda_X X^4 - \frac{1}{2} m_{HX} X |H|^2
\bigg).
\end{align}
Even at this level we may see the first qualitative difference between the flat and curved spacetime, namely in the flat spacetimes
the coefficient $a_{H0}$ vanishes which can be seen from the first two lines of (\ref{Xcl_cur}). This implies that the term 
proportional to $X^4$ will not contribute any operators of dimension six or less. Meanwhile, in curved spacetime the presence
of a nonzero $a_{H0}$ means that the term $X^4$ will introduce into the effective action new operators for the light field
with curvature dependent coefficients, we will call such operators gravity induced. These contributions 
(up to $dim O = 6,~\mathcal{R}^2$) are given by
\begin{align}
X|H|^2 &\approx - \frac{1}{m_{X}^2} \bigg \{
m_{HX} R |H|^2  + \nonumber \\
&+  \bigg [ m_{HX} + 2 \xi_X R \frac{m_{HX}}{m_{X}^2} + 4 \xi_X^2 R^2 \frac{m_{HX}}{m_{X}^4}
+2 \xi_X \frac{m_{HX}}{m_{X}^4} \square R \bigg ] |H|^4  + \nonumber \\
&+ \bigg [
\lambda_{HX}\frac{m_{HX}}{m_{X}^2} + 4 \xi_X \lambda_{HX} R \frac{m_{HX}}{m_{X}^4} 
\bigg ] |H|^6  + \nonumber \\
&+
\bigg [
\frac{m_{HX}}{m_{X}^2} + 2\xi_X R \frac{m_{HX}}{m_{X}^4}
\bigg ] |H|^2 \square |H|^2 + 4 \xi_X \frac{m_{HX}}{m_{X}^4} |H|^2 \square \left ( R |H|^2 \right )
\bigg \}, \\
X^3 &\approx - \frac{1}{m_{X}^6} \bigg \{
3 m_{HX}^3 R^2 |H|^2 + 3 \bigg [ m_{HX}^3 R + 4 \frac{m_{HX}^3 \xi_X}{m_{X}^2} R^2 
+ \frac{\lambda_{HX} m_{HX}^3}{m_{X}^2} R^2 \bigg ] |H|^4 + \nonumber \\
&+ \bigg [
m_{HX}^3 + 6 \frac{m_{HX}^3 \xi_X}{m_{X}^4} R + 24 \frac{m_{HX}^3 \xi_X^3}{m_{X}^4} R^2 + 
6 \frac{m_{HX}^3 \xi_x}{m_{X}^4} \square R + \nonumber \\
&+6 \bigg (
\frac{3}{2}\frac{\lambda_{HX} m_{HX}^3}{m_{X}^2}R + 6 \frac{\lambda_{HX} m_{HX}^3 \xi_X}{m_{X}^4}R^2
\bigg )
\bigg ] |H|^6 + \nonumber \\
&+ 6 \bigg [
\frac{ m_{HX}^3}{m_{X}^2}R + 6 \frac{ m_{HX}^3 \xi_X}{m_{X}^4}R^2
+ \frac{1}{2} \frac{\lambda_{HX} m_{HX}^3}{m_{X}^4}R^2  \bigg ] |H|^2 \square |H|^2 
\bigg \} \\
X^4 &\approx \frac{1}{m_{X}^8} \bigg \{
6 m_{HX}^4 R^2 |H|^4 +
\bigg [
4 m_{HX}^4 R + 16 \xi_X R^2 \frac{m_{HX}}{m_{X}^2} + 12 \lambda_{HX} \frac{m_{HX}^4}{m_{X}^2} R^2
\bigg ] |H|^6 + \nonumber \\
&+ 12 \frac{m_{HX}^4}{m_{X}^2} R^2 |H|^2 \square |H|^2
\bigg \}.
\end{align}

Before we present the results coming from the one-loop terms originated from integrating heavy fields let us make another 
simplification, namely from now on we put $\lambda_{X}=0$. This will make the presentation of the results more clear.
As a side note, we point out that keeping $\lambda_{X} \neq 0$  would result in an appearance of terms proportional to the powers 
of the $X$ field that should be replaced by (\ref{Xcl_cur}). This would give us additional contributions for the operators presented above
which would be suppressed by the numerical factor $\frac{\hbar}{64 \pi^2}$ and an appropriate power of $\lambda_X$.

To extract information about operators of dimension six that could be generated in the effective theory we will need the
HDW coefficients as defined by (\ref{coef_a0})-(\ref{coef_a3}). The two basic auxiliary quantities needed in the calculation 
are $Q$ and $P$ and are defined in (\ref{auxiliary_q_0})-(\ref{auxiliary_q_3}). 
The commutator of the covariant derivatives $W_{\alpha \beta}$ vanishes since the heavy field is an uncharged scalar. Taking this into account we may write
\begin{align}
\label{scalar_PQ}
Q = - \lambda_{HX} |H|^2 - 2 \xi_X R, \\
P = - \lambda_{HX}|H|^2 - \left ( 2 \xi_X - \frac{1}{6} \right )R.
\end{align}
In what follows we will skip operators of dimension greater then six, operators with coefficients containing curvature scalars 
of power three or higher and purely gravitational terms, all this will be symbolized by the use of the symbol '$\approx$'.
Moreover, from now on we will impose the $Z_2$ symmetry on the $X$ fields which leads to the condition $m_{3X} = 0$.
This symmetry implies that $X$ could represent the dark matter candidate.
Having this in mind we may write
\begin{align}
&P^3 \approx - \lambda_{HX}^3 |H|^6 - 3 \left ( 2 \xi_X - \frac{1}{6} \right ) \lambda_{HX}^2 R |H|^4 
+ \lambda_{HX} \left ( 2 \xi_X - \frac{1}{6} \right )^2 R^2 |H|^2, \\
&PZ_{(2)} + \frac{1}{2} B^{\mu} Z_{\mu} + \frac{1}{10} Z_{(4)} \approx 
- \frac{1}{2} \lambda_{HX}^2 d_{\mu} |H|^2 d^{\mu} |H|^2 
+ \lambda_{HX} \left (2 \xi_X - \frac{1}{30}  \right ) \square R |H|^2 + \nonumber \\
&- \frac{\lambda_{HX}}{10} \square^2 |H|^2 
- \frac{\lambda_{HX}}{30} \left ( \mathcal{K} - R_{\mu \nu} R^{\mu \nu} \right ) |H|^2
-\frac{\lambda_{HX}}{15} G^{\mu \nu} d_{\mu} d_{\nu} |H|^2,
\end{align} 
where $\mathcal{K} \equiv R_{\nu \nu \rho \sigma} R^{\nu \nu \rho \sigma} $ is the Kretschmann scalar and 
$G_{\mu \nu} \equiv R_{\mu \nu} - \frac{1}{2} g_{\mu \nu} R$ is the Einstein tensor. In the formulas above
we used integration by parts wherever necessary. This is possible since what enters into the effective action is not the 
HDW coefficients but their spacetime integrals. 
Below we present the result for the case when $X$ is $Z_{2}$ symmetric, which implies only quartic interaction among the scalars.
\begin{align}
S^{a_3}_{cEFT} &= \int \sqrt{-g} d^4 x \frac{\hbar}{(4 \pi)^2} \bigg \{
- \frac{\lambda_{HX}^2}{12 m_{X}^2} \frac{1}{2} d_{\mu} |H|^2 d^{\mu}|H|^2 - \frac{\lambda_{HX}^3}{12 m_{X}^2}|H|^6
+ \nonumber \\
&+ \bigg [
\frac{\lambda_{HX}}{12 m_{X}^2} \left ( 2 \xi_X - \frac{1}{6} \right )^2 R^2 + \frac{\lambda_{HX}}{4 m_{X}^2}
\left ( 2\xi_X - \frac{1}{30} \right ) \square R 
- \frac{\lambda_{HX}}{270 m_{X}^2} \left ( \mathcal{K} - R_{\mu \nu} R^{\mu \nu} \right ) 
\bigg ] |H|^2 + \nonumber \\
& - \frac{\lambda_{HX}^2}{4 m_{X}^2} \left ( 2 \xi_X - \frac{1}{6} \right ) R |H|^4
- \frac{\lambda_{HX}}{120 m_{X}^2} G^{\mu \nu} d_{\mu} d_{\nu} |H|^2
\bigg \},
\end{align} 
where $S^{a_3}$ denotes the part of the effective action that comes from the $a_3$ Hadamard-DeWitt coefficient, 
$S^{a_3} = \int \sqrt{-g} d^4 x \frac{\hbar}{64 \pi^2 } Tr \left ( \frac{ a_3}{3 m_X^2} \right )$, and
we included terms that are proportional to the tree-level operators in redefinitions of appropriate constants.  
As a check of our results we compared the coefficients of the first two operators to the flat spacetime case, see for example \cite{Henning_Lu_Murayama_2016}.
We found out that they are exactly the same, as expected. The operators in the second line represent the gravity induced contributions to the
Higgs mass parameter. Although they are expected to be small there is an interesting possibility that they may introduce a spacetime dependent
contribution to the critical temperature of the phase transition.
The last line contains the gravity induced contribution to the Higgs quartic coupling which will have its impact on the problem of the 
vacuum stability in the Standard Model, especially in the context of the early Universe.
In this line there is also an operator that couples the Higgs field kinetic term to the Einstein tensor. This last operator is actually
irrelevant for the dynamics of the Higgs fields since its contribution to the equations of motion vanishes due to the
vanishing of the four-divergence of the Einstein tensor.
As far as the $a_{4}$ term is concerned, most operators coming from it are of the order eight or higher (or are subleading contributions
to the already present ones). 
The relevant part that can contribute operators of the dimension up to six and terms up to second order in curvature invariants or
fourth metric derivatives is given by
\begin{align}
a_4 \approx P^4 + \frac{3}{5} \left \{ P^2 , Z_{(2)} \right \} + \frac{4}{5} P Z_{(2)}P + 
\frac{1}{5} \left \{P, Z_{(4)} \right \},
\end{align}
where all quantities were defined in (\ref{auxiliary_q_0})-(\ref{auxiliary_q_3}) and should be calculated with taking into account 
(\ref{scalar_PQ}), $h_{\mu} = 0$, $W_{\alpha \beta} = 0$ and discarding terms of the order $O(\mathcal{R}^3)$ or higher.
From the above formula for $a_4$ we obtained operators that contribute to the Higgs quartic coupling and the kinetic term.
After some algebra they are given by
\begin{align}
S^{a_{4}}_{cEFT} &= \int \sqrt{-g} d^4 x \frac{\hbar}{(4 \pi)^2} \frac{1}{48 m_{X}^4} \bigg \{
- \lambda_{HX}^2 \left ( 2 \xi_X - \frac{1}{10} \right ) R d_{\mu} |H|^2 d^{\mu} |H|^2 + \nonumber \\
&+ \frac{2}{15} \lambda_{HX}^2 G^{\mu \nu} d_{\mu} |H|^2 d_{\nu} |H|^2 + 
 \bigg [ 
6 \left ( 2 \xi_X - \frac{1}{6} \right )^2 \lambda_{HX}^2 R^2 + 
\frac{1}{15} \lambda_{HX}^2 \left ( \mathcal{K} - R_{\mu \nu} R^{\mu \nu} \right ) + \nonumber \\ 
&- \lambda_{HX}^2 \left ( - 12 \xi_X + \frac{108}{90} \right ) \square R 
+ \frac{8}{15} \lambda_{HX}^2 \nabla_{\mu} \nabla_{\nu} R^{\mu \nu}
\bigg ] |H|^4 +
4 \lambda_{HX}^3 \left ( 2 \xi_X - \frac{1}{6} \right ) R |H|^6
\bigg \}.
\end{align}

To sum up this section, let us write the cEFT for the Higgs doublet after integrating out the heavy $Z_{2}$-symmetric real
scalar singlet (the UV action is given by the $Z_2$ symmetric part of (\ref{actionUV_XH}))
\begin{align}
S_{cEFT} = \int \sqrt{-g} d^4 x \bigg(
&- \frac{1}{2} d^{\mu} H^{\dagger} d^{\mu} H -  \frac{1}{2} c_{dHdH} d_{\mu}|H|^2 d^{\mu}|H|^2
- c_{GdHdH} G^{\mu \nu} d_{\mu} |H|^2 d_{\nu}|H|^2 + \nonumber \\ 
&-\frac{1}{2}m_{H}^2 |H|^2 - \xi_X R |H|^2 - c_{H} |H|^2 - \frac{\lambda_{H}}{4!}|H|^4 - c_{HH}|H|^4 - c_{6}|H|^6
\bigg),
\end{align}
where we defined the curvature dependent coefficients in the following manner:
\begin{align}
\label{cdHdH}
c_{dHdH} &=   \frac{\hbar}{(4 \pi)^2} \frac{\lambda_{HX}^2}{12 m_{X}^2} 
\left (1 + \frac{\left ( \xi_X - \frac{1}{10} \right )}{m_{X}^2} R  \right ), \\ 
\label{GdHdH}
c_{GdHdH} &=  - \frac{\hbar}{(4 \pi)^2} \frac{\lambda_{HX}^2}{360 m_{X}^4}, \\
\label{cH}
c_{H} &=  \frac{\hbar}{(4 \pi)^2} \bigg [ \frac{\lambda_{HX}}{12 m_{X}^2} \left ( 2 \xi_X - \frac{1}{6} \right )^2 R^2 + \nonumber \\
&+ \frac{\lambda_{HX}}{4 m_{X}^2}
\left ( 2\xi_X - \frac{1}{30} \right ) \square R 
- \frac{\lambda_{HX}}{270 m_{X}^2} \left ( \mathcal{K} - R_{\mu \nu} R^{\mu \nu} \right ) \bigg ], \\
\label{cHH}
c_{HH} &= \frac{\hbar}{(4 \pi)^2} \bigg [
\frac{\lambda_{HX}^2}{4 m_{X}^2} \left ( 2 \xi_X - \frac{1}{6} \right ) R 
- \frac{\lambda_{HX}^2}{8 m_X^4} \left ( 2 \xi_X - \frac{1}{6} \right )^2  R^2  + \nonumber \\ 
&- \frac{\lambda_{HX}^2}{720 m_{X}^4} \left ( \mathcal{K} - R_{\mu \nu} R^{\mu \nu} \right ) 
+ \frac{\lambda_{HX}^2}{m_{X}^4} \left ( - \frac{1}{4} \xi_X + \frac{1}{40} \right ) \square R 
- \frac{\lambda_{HX}^2}{90 m_{X}^4}  \nabla_{\mu} \nabla_{\nu} R^{\mu \nu} 
\bigg ], \\
\label{c6}
c_6 &=   \frac{\hbar}{(4 \pi)^2} \frac{\lambda_{HX}^3}{12 m_{X}^2}
\left (1 - \frac{\left ( 2 \xi_X - \frac{1}{6} \right )}{m_{X}^2} R \right ).
\end{align}
In what follows we would like to comment on the revealed nature of the gravity induced 
contributions to the obtained effective field theory. We see that we have a linear in 
curvature contributions to the dimension six kinetic operator for the Higgs field, they are 
given by a part of the $c_{dHdH}$ coefficient and the $c_{GdHdH}$ one. Moreover, the
$c_{GdHdH} G^{\mu \nu} d_{\mu} |H|^2 d_{\nu}|H|^2$ term looks similar to the one named the non-minimal derivative coupling 
\cite{AMENDOLA1993175,PhysRevLett.105.011302,Nozari_Rashidi_2016}
that was analyzed in the context of the Higgs inflation. The difference is in the dimensionality of the operator,
the usual one is of dimension four $c_{GdHdH} G^{\mu \nu} d_{\mu} H^{\dagger} d_{\nu} H$, while the one obtained by us
is of dimension six. At this point it is worthy to note that our calculations indicate that the  
coupling of the Einstein tensor to the dimension four kinetic type operator does not arise after integration of 
the heavy scalar field. This implies that if the presence of such an operator could be inferred from the inflationary 
data it must be a remnant of the coupling of the Higgs field to the heavy field of a different statistic 
than a scalar field. From the structure of the HDW coefficients we may infer that
this probably will be a fermionic field, although the proof of this statement would demand calculations 
that are out of scope of this article.
 
The presence of the $c_H$ terms indicates that the Higgs mass parameter gets a contribution also from 
terms that are proportional to terms of order two in curvatures. This is hardly surprising
yet it nicely represents the general feature of the effective field theory in curved spacetime.
Namely, every effective operator present in the flat spacetime case will obtain contributions from 
terms proportional to the higher order curvature scalars or tensors. This means that the effective field 
theory in curved spacetime will be given by the action that represents expansion in both dimensions of the operators and 
powers of curvature invariants.

Before we turn to an analysis of the $c_{HH}$ coefficient we want to make a comment about the region of validity of our expansion 
in curvature invariants. Generally speaking, (\ref{Gamma_ren}) represents a valid contribution to the effective 
action for the light field if terms proportional to the higher order Hadamard-DeWitt coefficients present
in (\ref{Gamma_nr}) are dropped out. This implies that the operators of dimension eight should give smaller contributions
than these of dimension six and that $\frac{O(\mathcal{R}^5)}{m_{X}^6} << \frac{O(\mathcal{R}^4)}{m_{X}^4}$, 
where $O(\mathcal{R}^n)$ represents all curvature invariants of the order $n$, for example for $n=2$ we have
$O(\mathcal{R}^2) = \left \{ R^2, R_{\mu \nu} R^{\mu \nu}, \mathcal{K} \right \}$. Since at each order we have
new invariants that could not be expressed as powers of invariants from lower orders to determine the region of validity of our 
approximation we will slightly abuse the notation introduced above. 
From the relation $\frac{O(\mathcal{R}^5)}{m_{X}^6} << \frac{O(\mathcal{R}^4)}{m_{X}^4}$ we may infer that we have
$\frac{O(\mathcal{R})}{m_{X}^2} << 1$ and $\frac{O(\mathcal{R}^2)}{m_{X}^4} << 1$. Since we work only with terms that
are at most quadratic in curvature scalars, the last expression is enough to determine the maximal curvature allowed
to be analyzed by our approximation.

Now, let us return to the $c_{HH}$ coefficient. Firstly, let us note that in usual applications
the rate of change of the curvature is small, therefore we may disregard the last two terms in (\ref{cHH}).
Among the terms proportional to $O(\mathcal{R}^2)$ we have three possible hierarchies. 

The first one is when $R = 0$, which is the case for the vacuum solution to the Einstein equations, {\it{i.e.}}, Schwarzschild or Kerr 
black holes or for the radiation dominated Friedmann--Lema{\^i}tre--Robertson--Walker (FLRW) universe. In these situations 
the dominant contribution comes from the term proportional to $\mathcal{K} - R_{\mu \nu} R^{\mu \nu}$. 
\begin{figure}[tbp]
\centering
\includegraphics[width=.8\textwidth]{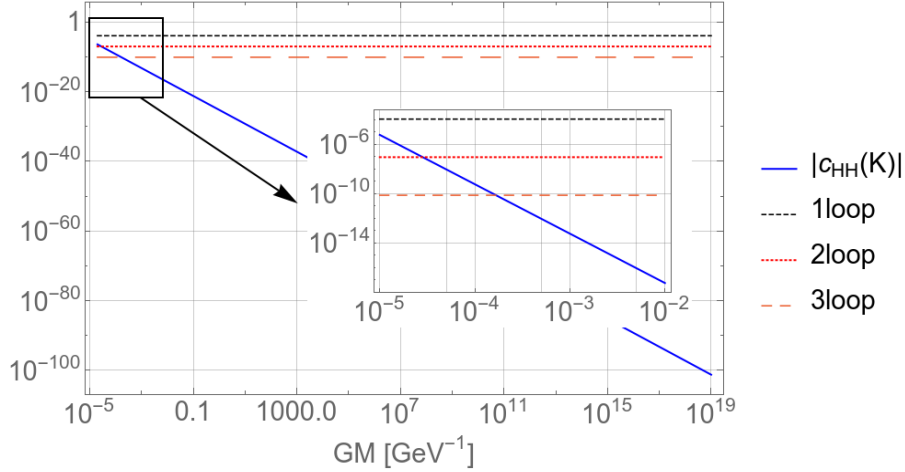}
\caption{The contribution of the gravity induced part of the $c_{HH}$ coefficient to the Higgs quartic coupling
in the black hole background. $|c_{HH}(K)| = |- \frac{1}{(4 \pi)^2} \frac{\lambda_{HX}^2}{720} \frac{\mathcal{K}}{m_{X}^4}|$,
$G$ is the Newton constant, $M$ is the black hole mass and loops prefactors are given by the formula 
nloop$= \frac{\lambda_{H}^{n+1}}{(16 \pi^2)^n}$. For the plot we chose $\lambda_{HX} = 0.25$, $\lambda_{H} = 0.13$
and $m_{X} = 10 {\rm TeV}$. The plot was made in the double logarithmic scale.
}
\label{fig1}
\end{figure}    
In Figure \ref{fig1} we plotted the contribution of the gravity induced operators given by $c_{HH}$ 
to the Higgs quartic coupling. The background spacetime was given by the Schwarzschild black hole for which the
relevant part of $c_{HH}$ is given by $c_{HH}(K) = - \frac{1}{(4 \pi)^2} \frac{\lambda_{HX}^2}{720} \frac{\mathcal{K}}{m_{X}^4}$.
The minimum mass of the black hole (maximum curvature of spacetime) that we can cover in our approximation
was estimated according to the following formula: $\frac{\mathcal{K}}{m_{X}^4} << 1$, 
where the Kretschmann scalar for the Schwarzschild black hole is given by $\mathcal{K} = \frac{48 (G M)^2}{r^6}$.
We calculated $\mathcal{K}$ at the innermost stable circular orbit which for the considered black hole is at $r = 6 GM$,
this gives us $\mathcal{K} = \frac{1}{972 (GM)^4}$. From this we get a rough estimate for the allowed mass, 
$GM \geq 10^{-7} {\rm GeV^{-1}}$. As an additional check of the validity of this formula we plugged 
the obtained estimation of $GM$ into the formula for the Hawking temperature for the Schwarzschild black hole ($T_{BH}$)
and we obtained $T_{BH} \sim 10^5 {\rm GeV}$. Since this temperature is bigger than the mass of the heavy particle, we
refined our estimate for the lower bound of $GM$ to be $GM \geq 10^{-5} {\rm GeV^{-1}}$, which corresponds to the 
temperature one order of magnitude smaller than the mass of the heavy particle. From the astrophysical perspective
the minimal allowed mass lays within the range of the allowed masses for the Primordial Black Holes (PBH) 
\cite{Georg2017} 
and corresponds roughly to the $M_{PBH} \sim 10^{10} {\rm g}$. The upper bound for $GM$ is not restricted in our
approximation. For the purpose of the plot we take it to be equal to the solar mass black hole $M_{BH} = M_{\odot}$.
The dotted and dashed lines represent the order of magnitude estimate for the 1, 2 and 3 loops self contribution
to the Higgs quartic coupling. They allow us to gauge the influence of the gravity induced term on the 
aforementioned Higgs coupling. As we may see from Figure \ref{fig1} the gravity contributions are 
irrelevant for large black holes, which is as expected. Meanwhile, for a small PBH, yet big enough not to 
evaporate due to the Hawking radiation up to the present time, the 
gravity induced contribution may be of the same order as the two loops effect. This implies that they
may be relevant for the vacuum stability around such a black hole. At this point let us note that
the current state of the art calculations pertaining to the stability of the Higgs vacuum in the flat spacetime
take into account at least some of the three loops effects.   
As to the nature of the contribution given by (\ref{cHH}) to the vacuum stability we may see that
since the relevant term has an opposite sign to the $-\frac{\lambda}{4!} |H|^4$ term in the Higgs potential 
it will lead to further instability of the vacuum in the vicinity of the black hole. 
As a final remark let us state that since we expect the PBH formed in the
early Universe to have even smaller masses, the obtained results indicate that further development of the
curved spacetime approach to the effective field theory may be instrumental in better understanding
of the influence of strongly gravitating objects on particle physics phenomena.  
\begin{figure}[tbp]
\centering
\includegraphics[width=.9\textwidth]{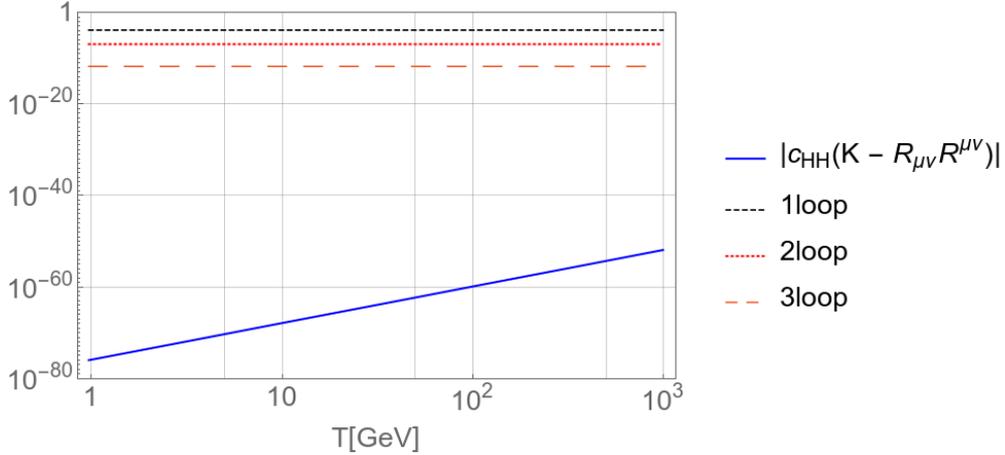}
\caption{The contribution of the gravity induced part of the $c_{HH}$ coefficient to the Higgs quartic coupling
in the radiation dominated FLRW background. 
$|c_{HH}(K)| = |- \frac{1}{(4 \pi)^2} \frac{\lambda_{HX}^2}{720} \frac{\mathcal{K} - R_{\mu \nu} R^{\mu\nu}}{m_{X}^4}|$,
$T$ is the temperature and loops prefactors are given by the formula 
nloop$= \frac{\lambda_{H}^{n+1}}{(16 \pi^2)^n}$. For the plot we chose $\lambda_{HX} = 0.25$, $\lambda_{H} = 0.13$
and $m_{X} = 10 {\rm TeV}$. The plot was made in the double logarithmic scale.
}
\label{fig2}
\end{figure}    
  
In Figure \ref{fig2} we presented the second example for which $R=0$, namely the radiation dominated FLRW universe.
Here we may connect the curvature of spacetime to the energy density ($\rho$) using the Einstein equations and radiation as a source of
the energy-momentum tensor $ \mathcal{K} - R_{\mu \nu} R^{\mu \nu}  = \frac{4}{3} \bar{M}_{Pl}^{-4} \rho^2$,
where $\bar{M}_{Pl}^{-2} = 8 \pi G$. Having in mind this relation we may find the maximal energy density allowed by our approximation
to be $\rho \leq 10^{43} {\rm GeV^4}$. Using the Stefan-Boltzmann law to connect the energy density to temperature we obtain $T \leq 10^8 {\rm GeV}$. 
This maximally allowed temperature should again be corrected due to the fact that we work with the effective field theory and 
we wish to integrate out particles with masses $m_{X} = 10^4 {\rm GeV}$. Taking into account this fact we set the maximal temperature 
to be $T = 10^3 {\rm GeV}$. As we may see from Figure \ref{fig2}, the gravity induced contributions to the Higgs quartic
coupling are always many orders of magnitude below the scale of the estimated three loops effects and therefore are of no 
consequence for the problem of the vacuum stability. At this point we want to remark that this is the case in the effective field theory,
and in the full theory where we treat both heavy and Higgs fields on equal footings this is not necessarily the case.

\begin{figure}[tbp]
\centering
\includegraphics[width=.9\textwidth]{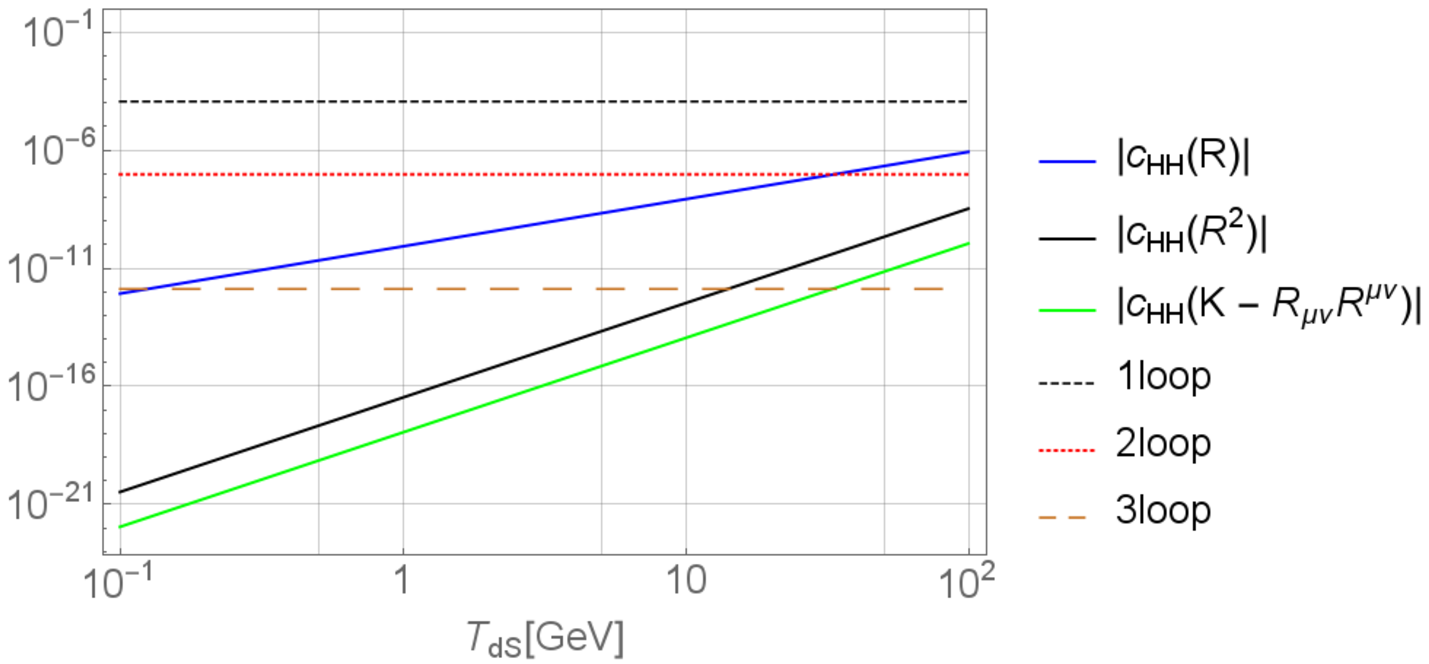}
\caption{The contribution of the gravity induced part of the $c_{HH}$ coefficient to the Higgs quartic coupling
in the de Sitter like FLRW background. The $c_{HH}$ is given by (\ref{cHH}). 
Loops prefactors are given by the formula nloop$= \frac{\lambda_{H}^{n+1}}{(16 \pi^2)^n}$ and 
$T_{dS}$ is the temperature of the de Sitter spacetime. 
For the plot we chose $\lambda_{HX} = 0.25$, $\lambda_{H} = 0.13$, $m_{X} = 10 {\rm TeV}$ and $\xi_X = 0$.
The plot was made in the double logarithmic scale.
}
\label{fig3}
\end{figure}

Now we will discuss the second hierarchy of terms in $c_{HH}$ for which $R \neq 0$ and $\xi \sim 0$. 
To illustrate our point we will use the de Sitter like stage of the FLRW universe. We may think of this 
as an FLRW universe filled with matter with the following equation of state: $p = -\rho$. Such a spacetime geometry
may be used to describe a part of the inflationary era of our Universe. In this case all terms of $c_{HH}$ contribute
(as earlier we disregard terms containing higher derivatives of the curvatures), the results are plotted 
in Figure \ref{fig3}.
To obtain the maximal allowed energy density we go through the same steps as in the $R = 0$ case. 
In the next step we translated the energy density to the temperature using the following formula: 
$T_{dS} = \frac{\sqrt{\Lambda}}{2 \sqrt{3} \pi}$ (see for example \cite{Parker_Toms_2009}),
where $\Lambda$ is the cosmological constant. To connect $\Lambda$ to the energy density
we used the Einstein equations in the FLRW background and the equation of state for matter mentioned earlier.
This resulted in the formula $T_{dS} = \frac{\bar{M}_{Pl}^{-1} \sqrt{\rho}}{2 \sqrt{3} \pi}$.
Due to the peculiarity of the de Sitter spacetime our effective field theory is valid in the whole 
range of energy density (temperature) allowed by the demand $\frac{O(\mathcal{R}^2)}{m_{X}^2} << 1$.  
The first thing we may infer from Figure \ref{fig3} is the fact that the term linear in $R$ dominates the remaining terms in $c_{HH}$. 
The second thing is the fact that, contrary to the radiation dominated universe, the gravity induced
contributions to the Higgs quartic coupling reach the same order of magnitude as the
two loops effects for large temperature. This implies that in calculations going beyond the one loop approximation we should
account at least for effective operators proportional to the Ricci scalar. 

\begin{figure}[tbp]
\centering
\includegraphics[width=.9\textwidth]{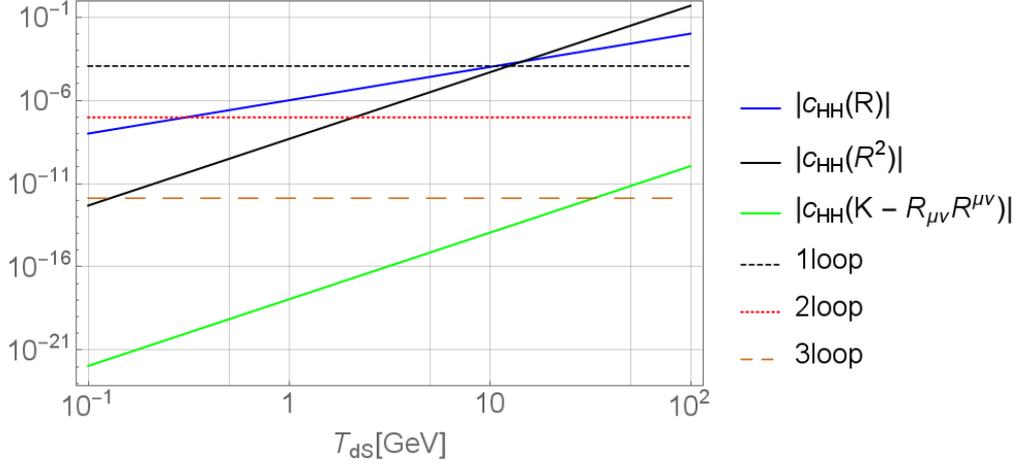}
\caption{The contribution of the gravity induced part of the $c_{HH}$ coefficient to the Higgs quartic coupling
in the de Sitter like FLRW background. The $c_{HH}$ is given by (\ref{cHH}). 
Loops prefactors are given by the formula nloop$= \frac{\lambda_{H}^{n+1}}{(16 \pi^2)^n}$ and 
$T_{dS}$ is the temperature of the de Sitter spacetime. 
For the plot we chose $\lambda_{HX} = 0.25$, $\lambda_{H} = 0.13$, $m_{X} = 10 {\rm TeV}$ and $\xi_X = 10^3$.
The plot was made in the double logarithmic scale.
}
\label{fig4}
\end{figure}

\begin{figure}[tbp]
\centering
\includegraphics[width=.9\textwidth]{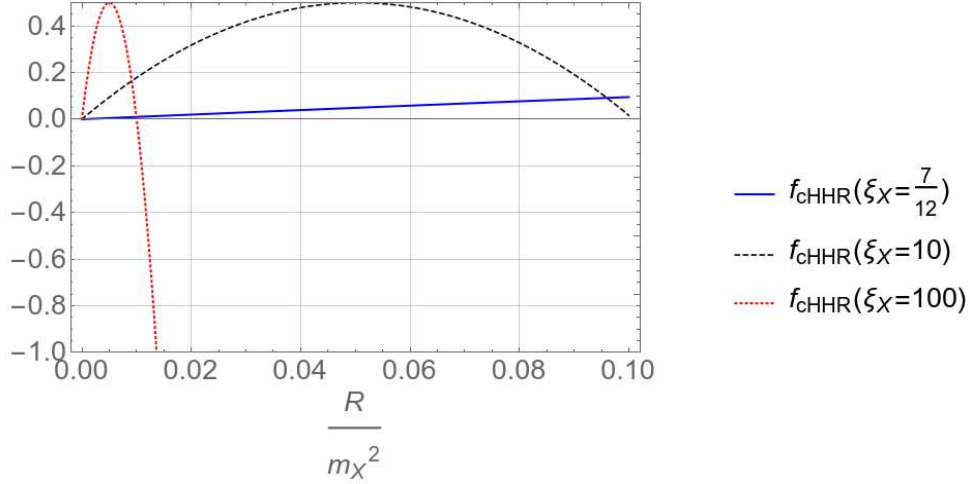}
\caption{The function $f_{cHHR} = \left (2 \xi_X - \frac{1}{6} \right ) \frac{R}{m_{X}^2}
- \frac{1}{2} \left (2 \xi_X - \frac{1}{6} \right )^2 \left ( \frac{R}{m_{X}^2} \right )^2$ for 
some values of the non-minimal coupling $\xi_X$. $m_{X}$ was held fixed at $10 {\rm TeV}$.
The maximally allowed curvature fulfills $\frac{R}{m_{X}^2} << 1$.
}
\label{fig5}
\end{figure}

The third type of hierarchy is for $R \neq 0$ and $\xi_{X} >> 1$ case. Again, as a background spacetime
we will take the de Sitter like phase of the FLRW universe. As far as the large non-minimal coupling to the spacetime curvature ($\xi_X$)
of the heavy scalar is concerned, it could be allowed to go up to $\xi_{max} \sim 10^{15}$ \cite{Atkins_Calmet_2013}.
The obtained results are plotted in Figure \ref{fig4}.
From Figure \ref{fig4} we may see that contributions to $c_{HH}$ from terms proportional to $\xi_X$
dominate over these coming form the $O(\mathcal{K} - R_{\mu \nu} R^{\mu \nu})$ term. Moreover, this term may be relevant
only when temperature is high enough and we are interested in calculations going beyond the two loops order. 
As far as the terms proportional to the Ricci scalar are concerned, the situation is quite different.
In the temperature range up to $T_{dS} \sim 20 {\rm GeV}$ the term linear in $R$ dominates,
while above this temperature the $R^2$ one gives a bigger contribution to $c_{HH}$.
The nature of this behavior could be inferred from the structure of these terms in the $c_{HH}$
coefficient, namely the structure at hand is $f_{cHHR} \equiv \left (2 \xi_X - \frac{1}{6} \right ) \frac{R}{m_{X}^2}
- \frac{1}{2} \left (2 \xi_X - \frac{1}{6} \right )^2 \left ( \frac{R}{m_{X}^2} \right )^2 $,
where we skipped the overall common factor. The behavior of this function with respect to the 
change of the spacetime Ricci scalar $\frac{R}{m_{X}^2}$ is plotted in Figure \ref{fig5}.
As we may see from it, $f_{cHHR}$ becomes negative for large enough $R$ (large temperature) which implies
that terms proportional to $R^2$ dominate over the one linear in $R$. This dominance is not observed for small
$\xi_X$ because it happens at the value of $\frac{R}{m_{X}^2}$ that is beyond the range of validity of our approximation.  
On this example we may see an additional subtlety that becomes apparent when the large non-minimal coupling to gravity is 
considered. Namely, in this case the validity of our approximation in calculation of the form of effective operators coming from
loops of heavy fields needs to modify previous formula for the maximally allowed spacetime curvature 
to be $\frac{( 2 \xi_X -\frac{1}{6}) R}{m_{X}^2} << 1$, or for sufficiently big $\xi_X$ $\frac{\xi_X R}{m_{X}^2} << 1$.
This last formula gives us either a more stringed constraint on the allowed spacetime curvature or on the maximal 
value of $\xi_X$ that can be covered by our effective field theory.    
\begin{figure}[tbp]
\centering
\includegraphics[width=.9\textwidth]{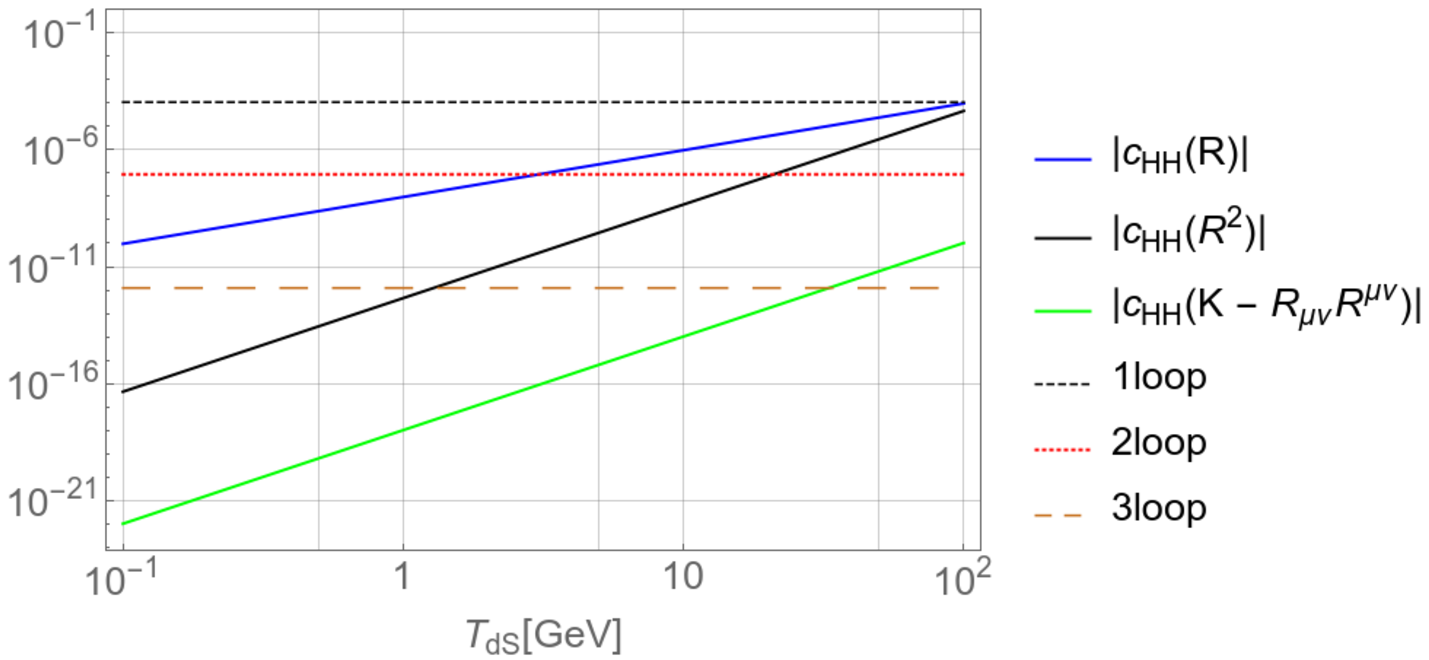}
\caption{The contribution of the gravity induced part of the $c_{HH}$ coefficient to the Higgs qaurtic coupling
in the de Sitter like FLRW background. The $c_{HH}$ is given by (\ref{cHH}). 
Loops prefactors are given by the formula nloop$= \frac{\lambda_{H}^{n+1}}{(16 \pi^2)^n}$ and 
$T_{dS}$ is the temperature of the de Sitter spacetime. 
For the plot we chose $\lambda_{HX} = 0.25$, $\lambda_{H} = 0.13$, $m_{X} = 10 {\rm TeV}$ and $\xi_X = 10$.
The plot was made in the double logarithmic scale.
}
\label{fig6}
\end{figure}    
Figure \ref{fig6} represents the same type of a plot as Figure \ref{fig5} but for $\xi_X =10$.
We see that the term linear in $R$ dominates contributions to $c_{HH}$ in the whole range of allowed temperatures. 
Moreover, we see that for sufficiently high temperature, for the displayed parameter it is roughly above $T_{dS} \sim 50 {\rm GeV}$
the gravity induced operators will contribute to the effective Higgs quartic coupling on the same level like the one-loop effects. 
As a final note let us point out that the term linear in $R$ in $c_{HH}$ has the same sign as $\lambda_{H}$,
therefore gravity leads to improvement of the vacuum stability in the de Sitter spacetime. 
  
The last coefficient that remains to be discussed is $c_6$. Here we see that taking into account the
presence of the spacetime curvature leads to its slight decrease as compared to the flat spacetime case.

\subsection{Yukawa model with the heavy real scalar}
\label{subsec:yukawa}

In this subsection we will present a construction of the effective field theory in curved spacetime for Dirac fermions 
interacting with a heavy real scalar singlet. The UV action may be written as
\begin{align}
\label{actionUV_fX}
S_{UV} = \int \sqrt{-g} d^4 x \bigg(
&i \bar{\psi} \gamma_{\mu} d^{\mu} \psi - m \bar{\psi} \psi - y_{X} X \bar{\psi} \psi - \frac{1}{2} d_{\mu} X d^{\mu} X - \frac{1}{2} m_{X}^2 X^2 - \xi_X R X^2 
\bigg).
\end{align}
The classical equation of motion for the scalar is
\begin{align}
\bigg( \square - m_{X}^2  - 2 \xi_X R \bigg) X = y_X \bar{\psi} \psi.
\end{align}
From the above we get
\begin{align}
\label{psiXcl}
X_{cl} = \frac{1}{\square - m_{X}^2  - 2 \xi_X R }y_X \bar{\psi} \psi.
\end{align}
After expanding this in the powers of $m_{X}^{-2}$ we get a local approximation
\begin{align}
X_{cl} \approx  - \frac{1}{m_{X}^2} \bigg ( 1
+ \frac{\square   -  2 \xi_X R}{m_{X}^2} + 
\frac{\square   -  2 \xi_X R}{m_{X}^2} \frac{\square   - 2 \xi_X R}{m_{X}^2}
\bigg ) y_X \bar{\psi} \psi.
\end{align}
Plugging this back into (\ref{actionUV_fX}) and keeping only operators of dimension six and less we get
the cEFT for fermions in curved spacetime
\begin{align}
S_{cEFT} = \int \sqrt{-g} d^4 x \bigg \{
i \bar{\psi} \gamma_{\mu} d^{\mu} \psi - m \bar{\psi} \psi
+ c_{6} (\bar{\psi} \psi)^2
\bigg \},
\end{align}
where 
\begin{align}
\label{c6psi}
c_{6} =
 \frac{y_X^2}{m_{X}^2} \bigg (1 - \frac{2 \xi_X}{m_{X}^2} R 
+ \frac{4 \xi_X^2 }{m_{X}^4} R^2 - \frac{2 \xi_X }{m_{X}^4} \square R \bigg ).
\end{align}
The first observation is that $c_6$ contains only terms proportional to the Ricci scalar, but not to the 
other curvature scalars which stems from the fact that we work only with operators coming from (\ref{psiXcl}).
Although the heavy scalar loops do not contribute to the matter part of the effective field theory,
we need to consider them in finding the allowed range of spacetime curvatures.   
The second observation is that while in the flat spacetime case the presence of the dimension six operator leads 
to an appearance of the vacuum expectation value $<\bar{\psi} \psi> = \frac{m}{c_6}$, the presence of the gravity induced operators 
leads to diminishing of this vev, provided that the spacetime is the one of constant curvature $R = const.$ 
(like for example the de Sitter spacetime). On the other hand, if $R \neq const.$ we cannot determine    
the $<\bar{\psi} \psi>$ by simply solving an algebraic equation, instead we need to solve partial differential equations coming
from equations of motion.

\subsection{Quantum Electrodynamics with integrated out fermions}
\label{subsec:QED}

In this subsection we work out the QED example with the heavy fermionic sector.
The starting action for the matter sector is 
\begin{align}
S_{UV} = \int \sqrt{-g} d^4 x \bigg(
- \frac{1}{4} F_{\mu \nu} F^{\mu \nu} + i \bar{\psi} \gamma^{\mu} d_{\mu} \psi - m \bar{\psi} \psi
\bigg).
\end{align}
The $F_{\mu \nu} \equiv 2 \nabla_{[ \mu} A_{\nu ]}$ is the standard Maxwell tensor for the $U(1)$ gauge field $A_{\mu}$
and a covariant derivative for the fermionic field is given by $d_{\mu} \equiv \nabla_{\mu} - i e A_{\mu}$.
The second functional derivative of $S_{UV}$ with respect to the heavy fermionic field gives us the 
following operator:
\begin{align}
D_{-m} = i \gamma^{\mu} d_{\mu} - m.
\end{align}
To bring it to the form (\ref{operatorD2}) we will use the following formula: $\ln det D = \frac{1}{2} \ln det D^2$
and the fact that the aforementioned operator and $D_{+m} \equiv i \gamma^{\mu} d_{\mu} + m$ have the same spectrum of eigenvalues.
Alternatively, we may redefine the path integral variables, see for example \cite{Buchbinder_Odintsov_Shapiro_1992}.
After this operation we obtained
\begin{align}
\label{DiracD2}
D^2 = D_{-m} D_{+m} = - \gamma^{\mu} d_{\mu} \gamma^{\nu} d_{\nu} - m^2 = \square {\bf{1}} 
 -  \frac{1}{4} R {\bf{1}} + \frac{ie}{2} F_{\alpha \beta} \gamma^{\alpha} \gamma^{\beta} - m^2,
\end{align}
where ${\bf{1}}$ is a four-by-four unit matrix and we used the definition of gamma matrices in curved spacetime
$\{ \gamma^{\mu}, \gamma^{\nu} \} = -2 g^{\mu \nu} {\bf{1}}$ and the fact that 
$W_{\mu \nu} \psi \equiv [d_{\mu} , d_{\nu}] \psi = 
\left ( -  \frac{1}{4} R_{\mu \nu \alpha \beta} \gamma^{\alpha} \gamma^{\beta} - i e F_{\mu \nu} \right ) \psi$. 
Below we present some useful and well known properties of the trace of gamma matrices 
(these formulas take into account the chosen signature of metric tensor)
\begin{align}
&tr \gamma^{\mu} \gamma^{\nu} = -4 g^{\mu \nu}, \\
&tr \gamma^{\alpha} \gamma^{\beta} \gamma^{\mu} \gamma^{\nu}  = 4 \left (
g^{\alpha \beta} g^{ \mu \nu} - g^{\alpha \mu} g^{\beta \nu} + g^{\alpha \nu} g^{\beta \mu}
\right ), \\
&tr \gamma^{\alpha} \gamma^{\beta} \gamma^{\mu} \gamma^{\nu} \gamma^{\rho} \gamma^{\sigma} = 
4 \bigg [
- g^{\alpha \beta} \left ( 
g^{\mu \nu} g^{\rho \sigma} - g^{\mu \rho} g^{\nu \sigma} + g^{\mu \sigma} g^{\nu \rho}
\right )
+ g^{\alpha \mu} \left ( 
g^{\beta \nu} g^{\rho \sigma} - g^{\beta \rho} g^{\nu \sigma} + g^{\beta \sigma} g^{\nu \rho}
\right ) + \nonumber \\
&- g^{\alpha \nu} \left ( 
g^{\beta \mu} g^{\rho \sigma} - g^{\beta \rho} g^{\mu \sigma} + g^{\beta \sigma} g^{\mu \rho}
\right )
+ g^{\alpha \rho} \left ( 
g^{\beta \mu} g^{\nu \sigma} - g^{\beta \nu} g^{\mu \sigma} + g^{\beta \sigma} g^{\mu \nu}
\right ) + \nonumber \\
&- g^{\alpha \sigma} \left ( 
g^{\beta \mu} g^{\nu \rho} - g^{\beta \nu} g^{\mu \rho} + g^{\beta \rho} g^{\nu \mu}
\right )
\bigg ].
\end{align}
Comparing (\ref{DiracD2}) with (\ref{operatorD2}) we found (from now on we will skip writing the unit matrix {\bf{1}} 
to shorten the notation)
\begin{align}
&h^{\mu} = 0, \\
&Q = \Pi = - \frac{1}{4} R + \frac{i e}{2} F_{\alpha \beta} \gamma^{\alpha} \gamma^{\beta} ,\\
&P = - \frac{1}{12} R  + \frac{i e}{2} F_{\alpha \beta} \gamma^{\alpha} \gamma^{\beta}, \\
&J_{\mu} = d_{\alpha} W^{\alpha }_{~~\mu} = - \frac{1}{4} \nabla_{\alpha} R^{\alpha}_{~~ \mu \nu \rho} \gamma^{\nu} \gamma^{\rho}
- ie \nabla_{\alpha} F^{\alpha}_{~~ \mu}.
\end{align} 
The remaining quantities that need to be calculated are given below. We write only terms that give an operator of dimension six or less
and contain terms at most linear in curvatures and survive after taking a trace with respect to gamma matrices.
\begin{align}
&tr \left ( B_{\mu} Z^{\mu} \right ) = tr \left (d_{\mu}P d^{\mu} P - \frac{1}{9} J_{\mu} J^{\mu} \right ) =
2 e^2 \nabla_{\mu} F_{\alpha \beta} \nabla^{\mu} F^{\alpha \beta}
+ \frac{4}{9} e^2 \nabla_{\alpha} F^{\alpha}_{~~ \mu} \nabla_{\beta} F^{\beta \mu}, \\
&tr P^3 = - \frac{1}{2} e^2 R F_{\alpha \beta} F^{\alpha \beta}, \\
&tr \left ( P \square Q \right ) = 2 e^2 F_{\alpha \beta} \square F^{\alpha \beta}, \\
&tr \left ( W_{\alpha \beta} W^{\alpha \beta} P \right ) = \frac{e^2}{3} R F_{\alpha \beta} F^{\alpha \beta} 
+ 2 e^2 F_{\alpha \beta} F_{\mu \nu} R^{\alpha \beta \mu \nu}, \\
&tr \left ( J_{\mu} J^{\mu} \right ) = - 4e^2 \nabla_{\alpha} F^{\alpha}_{~\mu} \nabla_{\beta} F^{\beta \mu}, \\
&tr \left ( W_{\alpha \beta} d^{\alpha} J^{\beta} \right ) = - 4 e^2 F_{\alpha \beta} \nabla^{\alpha} \nabla_{\mu} F^{\mu \beta},\\
&tr \left ( d_{\mu} W_{\alpha \beta} d^{\mu} W^{\alpha \beta} \right ) = -4e^2 \nabla_{\mu} F_{\alpha \beta} \nabla^{\mu} F^{\alpha \beta},\\
&tr \left ( R^{\alpha \beta} W^{\mu}_{~ \alpha} W_{\mu \beta} \right ) = - 4 e^2 R^{\alpha \beta} F^{\mu}_{~ \alpha} F_{\mu \beta}, \\
&tr \left ( R^{\alpha \beta \mu \nu} W_{\alpha \beta} W_{\mu \nu} \right ) = -4e^2 R^{\alpha \beta \mu \nu} F_{\alpha \beta} F_{\mu \nu},\\
&tr \left ( W_{\mu \nu} W^{\nu}_{~ \alpha} W^{\alpha \mu} \right ) = 0, \\
&tr Q_{(4)} = 0,     
\end{align}
the $tr \left ( W_{\mu \nu} W^{\nu}_{~ \alpha} W^{\alpha \mu} \right )$ is zero since in the end it can be written as 
a product of a symmetric and anti-symmetric tensors, for example $F_{\mu \nu} F^{\nu}_{~ \alpha} F^{\alpha \mu} = T_{\mu \alpha} F^{\alpha \mu}$,
where $T_{\mu \alpha} = F_{\mu \nu} F_{\rho \alpha} g^{\nu \rho} = T_{\alpha \mu}$.
Taking the above formulas into account we may write 
\begin{align}
\int \sqrt{-g} d^4 x sTr \left (  a_{3} \right ) &= \int \sqrt{-g} d^4 x  \left (  \frac{1}{2} \right ) \bigg(
- \frac{1}{3} e^2 R F_{\mu \nu} F^{\mu \nu} - \frac{4}{30} e^2 R_{\alpha \beta \mu \nu} F^{\alpha \beta} F^{\mu \nu}
+ \nonumber \\
&+ \frac{52}{30} e^2 R_{\alpha \beta} F^{\mu \alpha} F_{\mu}^{~ \beta} 
- \frac{144}{90} e^2 \nabla_{\mu} F^{\mu}_{~ \alpha} \nabla_{\nu} F^{\nu \alpha}
\bigg).
\end{align} 
The $ \frac{1}{2}$ factor comes form the fact that we worked with the operator $D^2$ and not with the Dirac operator $D_{-m}$ and 
to obtain the above formula we used Bianchi identities for the Maxwell field-strength tensor and the Riemann tensor
\begin{align}
\nabla_{\mu} F_{\alpha \beta} + \nabla_{\alpha} F_{\beta \mu} + \nabla_{\beta} F_{\mu \alpha} =0, \\
R_{\alpha \beta \mu \nu} + R_{\alpha \mu \nu \beta} + R_{\alpha \nu \beta \mu} = 0,
\end{align} 
and also the definition of the comutator of covariant derivatives acting on a tensor field
\begin{align}
\left [ \nabla_{\rho} , \nabla_{\sigma} \right ] X^{\mu}_{~ \nu}  = 
R^{\mu}_{~ \lambda \rho \sigma} X^{\lambda}_{~ \nu} - R^{\lambda}_{~ \nu \rho \sigma} X^{\mu}_{~ \lambda}. 
\end{align}
Considering the above formulas we end up with the following expression for the cEFT for the $U(1)$ vector field:
\begin{align}
S_{cEFT} &= \int \sqrt{-g} d^4 x \bigg[
- \frac{1}{4} F_{\mu \nu} F^{\mu \nu} - \frac{1}{m^2}  \frac{e^2 \hbar}{(4 \pi)^2} 
\bigg( 
- \frac{144}{1080} \nabla_{\mu} F^{\mu}_{~ \alpha} \nabla_{\nu} F^{\nu \alpha}
- \frac{1}{36} R F_{\mu \nu} F^{\mu \nu} 
+ \nonumber \\
&+ \frac{52}{360} R_{\alpha \beta} F^{\mu \alpha} F_{\mu}^{~ \beta} 
- \frac{4}{360}  R_{\alpha \beta \mu \nu} F^{\alpha \beta} F^{\mu \nu}
\bigg)
\bigg].
\end{align}
The corrections to the photon behavior stemming from the above effective action were already discussed in the literature 
in the early '80 \cite{PhysRevD.22.343} and more recently in \cite{Bastianelli_2009,Domenech_2018}, therefore the above example 
constitutes another check of the validity of our method of obtaining the effective field theory in curved spacetime.

\section{Summary}
\label{sec:summary}

In the presented article we checked if the heat kernel method could provide a viable tool in a systematic
construction of the effective field theory in curved spacetime. Our calculations confirmed that 
the aforementioned method, already used with successes in the construction of the one-loop effective action in 
curved spacetime, can be a valuable tool in building the curved spacetime effective field theory (cEFT).
Moreover, we want to point out that this approach can be viewed as a direct generalization of the universal
effective action method proposed recently for construction of the effective field theory in flat spacetime.

After describing the ingredients of the heat kernel method that are necessary in the task at hand we 
worked out three examples which allowed us to both explain in detail the required steps and check the validity of our approach. 
At this point let us remind that, as was pointed in the introduction, we worked
out only effects coming from the heavy-heavy loops. The effects of the heavy-light or
light-light loops may also be computed by this method but taking them into account in
the current work would lead to unnecessary computational complications that would 
dim the presentation of the method.

The first example considered was obtaining the curved spacetime effective field theory for the Higgs doublet
after integrating out the real heavy scalar singlet. In subsection \ref{subsec:singlet} we presented the resulting cEFT 
containing operators up to dimension six and the gravity induced coefficients containing terms up to second order in curvatures. 
As an immediate test of our calculations we compared the obtained coefficients to the flat spacetime ones presented in \cite{Henning_Lu_Murayama_2016}
and we found that they agreed.

In the next step we analyzed what new effects the gravity induced terms may possibly introduce in the case of a few chosen
physically interesting spacetime backgrounds. These backgrounds were the small mass black hole (in the mass range 
experimentally allowed for the Primordial Black Holes), the radiation dominated  
Friedmann--Lema{\^i}tre--Robertson--Walker universe and the FLRW universe in the de Sitter stage.

Firstly, we found out that integrating out the heavy scalar field will generate a non-minimal derivative coupling
of the Higgs field to gravity. An existence of such a coupling of dimension four Higgs kinetic operator was 
postulated in the context of one of the Higgs inflation models. On the other hand, what we found is that the non-minimal derivative 
coupling is between the Einstein tensor and the dimension six Higgs kinetic operator only.
We also observed that every operator present in the flat spacetime effective field theory obtains an infinite tower 
of contributions proportional to higher and higher powers of curvatures. Fortunately, they come with suppression
factors proportional to adequately high powers of an inverse of the heavy field mass. Therefore, for most cases it will be sufficient 
to consider only terms up to second order in curvatures. Taking into account terms proportional to the curvature squared
is important because for some interesting spacetimes like for example the Kerr black hole or the radiation dominated FLRW universe
the Ricci scalar vanishes identically.

Next, we turn our attention to an analysis of the influence of the gravity induced dimension four operators on the Higgs quartic 
selfcoupling. We found out that for the PBH with mass in the range of $10^{10} - 10^{11} {\rm g}$ modelled by the Schwarzschild metric
the contribution of the coefficient proportional to the Kretschmann scalar $\mathcal{K}$ may be bigger than the two-loops effects 
coming from Higgs quartic selfinteraction. The results were depicted in Figure \ref{fig1}.
As is evident from Figure \ref{fig2} the gravity induced terms are of no consequence for the 
Higgs quartic coupling if we choose the spacetime to be described by the energy dominated FLRW universe. 
Lastly, we analyzed the case when the spacetime is given by the de Sitter like metric (strictly speaking, it was the FLRW metric
for which matter possessed the following equation of state: $p = - \rho$). The obtained results were presented 
in Figures \ref{fig3} and \ref{fig6}. Such a spacetime may represent an end of the inflationary era 
just before reheating or if we model the reheating as a process that takes some time this metric should be 
also valid for at least a part of a timespan of reheating. 
In any case, it turned out that for the de Sitter metric sourced by a sufficiently high energy density, but within the
limit of validity of our approximation, the gravity induced coefficients may dominate the two-loops effects.

We also derived the gravity induced contribution to the coefficient of the dimension six operator $|H|^6$.
From the formula (\ref{c6}) we may see that its presence leads to a slight decrease of the value of this
coefficient as compared to the flat spacetime case. 

To generalize the results of this subsection we may formulate a few statements concerning the curved spacetime effective field theory. 
Firstly, the nature of the background spacetime, by which we mean vanishing (or not) of the Ricci scalar,
dictates whether the calculation should be done up to terms linear or quadratic in curvatures. 
Secondly, for the non-vanishing $R$ it is sufficient to keep only contributions of terms linear in curvatures
to the coefficients of the operators of dimension up to four. 
Thirdly, if we are interested in the cEFT containing operators of dimension six then
we should keep these gravity induced ones that are not present in flat spacetime, like 
the non-minimal derivative coupling but we may probably skip the gravity contribution to the ones that 
are already present in flat spacetime like $|H|^6$ in the Higgs case.

In subsection \ref{subsec:yukawa} we discussed integrating out the heavy scalar in the Yukawa model.
This example is somewhat simplistic, since it involves only finding a local approximation for the 
classical solution for the equation of motion of the heavy field. There are no contributions 
to the fermionic part of the effective theory coming from the scalar loops since the integrated out scalar does not possess
a selfinteraction term in the UV action. Despite this we still found that gravity contributes 
to the coefficient of the dimension six operator. This leads to a modification to the 
vacuum expectation value for the fermionic bilinear (in case $R = const.$). 
Moreover, although this modification seems to be trivial it is not so from the computational standpoint.
Namely, to find the fermionic field vev in the case $R \neq const.$  we need to solve partial differential 
equations and not an algebraic equation. As a side note, let us point out that the same is true for 
finding the vev of the Higgs field in the case when it is coupled non-minimally to gravity.

In the last subsection of section \ref{sec:examples} we rederived the effective field theory for 
photons in curved spacetime after integrating out fermions from QED. The obtained results were 
already known and discussed in the {'80}. Therefore, this subsection served more as a working 
example as how to integrate out fermionic field and as an additional check of validity of the obtained formulas.     

To conclude, the presented results indicate that the heat kernel method 
may be a viable way of extending the concept of the systematically obtained effective action to curved spacetime. 
Additionally, this type of cEFT may be vital in an analysis of particle physics phenomena in the strong
gravity regime. As interesting fields for further practical applications we want to point out
the problem of seeding vacuum instability by PBH, a question of the bariogenesis processes 
around such objects and an influence of the gravity induced operators on reheating or inflation.

\acknowledgments


{\L}N  was supported by the National Science Centre, Poland under a grant DEC-2017/26/D/ST2/00193.

\appendix
\section{The Hadamard-DeWitt coefficients used in the paper}
\label{appA}

Below we present a list of the Hadamard-DeWitt coefficients relevant to the 
problem of obtaining the curved spacetime effective action (cEFT).
The form of the coefficients and the notation follows closely \cite{Avramidi_2000},
with the exception of the name change for the commutator of the covariant derivatives
(in this article it is called $W_{\alpha \beta}$ while in \cite{Avramidi_2000} it is $\mathcal{R}_{\alpha \beta}$)
and $W_{\alpha \beta}$ defined in (2.153) on page 40 of \cite{Avramidi_2000} here was named as $\mathcal{W}_{\alpha \beta}$.   
Since in the presented article we concentrated on the cEFT containing terms at most of  the second 
order in curvatures or equivalently fourth derivatives of the metric and operators of the dimension 
up to six we will present $a_3$ and $a_4$ coefficients with this accuracy.  
The basic quantities that are needed for the construction of the coefficients are read off from the 
form of the operator given in (\ref{operatorD2}) and are 
\begin{align}
Q &= \Pi - d_{\mu}h^{\mu} - h_{\mu}h^{\mu}, \\
W_{\alpha \beta} &= [d_{\alpha}, d_{\beta}] - 2 d_{[ \alpha} h_{\beta ]} - 2 h_{[ \alpha} h_{\beta ]}, \\
J_{\mu} &= d_{\alpha} W^{\alpha}_{~\mu}, \\
Z_{\mu} &= d_{\mu} P - \frac{1}{3}J_{\mu}, \\
B_{\mu} &= d_{\mu} P + \frac{1}{3}J_{\mu}, \\
Z_{(2)} &= \square \left ( Q + \frac{1}{5} R \right ) +
 \frac{1}{30} \left ( R_{\alpha \beta \gamma \delta } R^{\alpha \beta \gamma \delta } - R_{\mu \nu} R^{\mu \nu} \right )
 + \frac{1}{2} W_{\alpha \beta} W^{\alpha \beta}, \\
 Z_{(4)} &= Q_{(4)} + 2 \left \{  W^{\mu \nu} , d_{\mu} J_{\nu} \right \} + \frac{8}{9} J_{\mu} J^{\mu}
 + \frac{4}{3} d_{\mu} W_{\alpha \beta} d^{\mu} W^{\alpha \beta}  + \nonumber \\
 &+ 6 W_{\mu \nu} W^{\nu }_{~~ \gamma} W^{\gamma \mu}
 + \frac{10}{3} R^{\alpha \beta} W^{\mu}_{~~\alpha} W_{\mu \beta} - R^{\mu \nu \alpha \beta}W_{\mu \nu} W_{\alpha \beta}
 + O(\mathcal{R}^3), \\
 Q_{(4)} &= \square^2 Q - \frac{1}{2} \left [ W^{\mu \nu}, \left [ W_{\mu \nu} ,Q \right ] \right ]
 - \frac{2}{3} \left [ J^{\mu}, d_{\mu} Q \right ] + \frac{2}{3} R^{\mu \nu} d_{\mu} d_{\nu}Q
 + \frac{1}{3} d_{\mu}R d^{\mu}Q.
\end{align} 
The first five coefficients are given by (the reader should remember that the
unit matrix $\mathbb{1}$ should be put wherever it is necessary to keep the correct dimension of the appropriate terms)
\begin{align}
a_0 &= 1, \\
a_1 &\equiv P  = Q  + \frac{1}{6}R , \\
a_2 &= P^2 + \frac{1}{3} Z_{(2)}, \\
a_3 &= P^3 + \frac{1}{2} \left \{ P, Z_{(2)} \right \} + \frac{1}{2} B^{\mu} Z_{\mu} + \frac{1}{10} Z_{(4)}, \\
a_4 &= P^4 + \frac{3}{5} \left \{ P^2 , Z_{(2)} \right \} + \frac{4}{5} P Z_{(2)}P + 
\frac{4}{5} \left\{P, B^{\mu} Z_{\mu}\right \} + \frac{2}{5} B^{\mu} P Z_{\mu} + \nonumber \\
&-\frac{2}{5} B_{\mu} Y^{\nu \mu} Z_{\mu} + \frac{1}{3} Z_{(2)} Z_{(2)} + \frac{2}{5} B^{\mu} Z_{\mu (2)}
+\frac{2}{5}C^{\mu} Z_{\mu} + \frac{1}{5} \left \{P, Z_{(4)} \right \} + \nonumber \\
&+ \frac{4}{15} \mathcal{D}^{\mu \nu} Z_{\mu \nu} + \frac{1}{35} Z_{(6)},
\end{align} 
where 
\begin{align}
Y_{\mu \nu} &= W_{\mu \nu} + \frac{1}{3}R_{\mu \nu}, \\
G_{\mu} &= - \frac{1}{5} \square J_{\mu} - \frac{2}{15} \left [ W_{\alpha \mu} , J^{\alpha} \right ]
- \frac{1}{10} \left [ W_{\alpha \beta}, d_{\mu} W^{\alpha \beta} \right ] 
- \frac{2}{15} R^{\alpha \beta} d_{\alpha} W_{\beta \mu} + \nonumber \\
&+ \frac{2}{15} R_{\mu \alpha \beta \gamma} d^{\alpha} W^{\beta \gamma} - \frac{7}{45} R_{\mu \alpha }J^{\alpha}
+ \frac{2}{5}d_{\alpha} R_{\beta \mu} W^{\beta \alpha} - \frac{1}{15} d^{\alpha}R W_{\alpha \mu}, \\
Q_{\mu (2)} &= d_{\mu} \square Q + \left [ W_{\nu \mu}, d^{\nu}Q \right ]
+ \frac{1}{3} \left [ J_{\mu} , Q \right ] + \frac{2}{3}R^{\nu}_{~\mu} d_{\nu}Q, \\
V_{\mu} &= Q_{\mu (2)} + \frac{1}{2} \left \{ W_{\alpha \beta}, d_{\mu} W^{\alpha \beta} \right \}
- \frac{1}{3} \left \{ J^{\nu}, W_{\nu \mu}\right \} + O(\nabla^5 g), \\
Z_{\mu (2)} &= V_{\mu} + G_{\mu}, \\
C_{\mu} &= V_{\mu} - G_{\mu}, \\
\mathcal{W}_{\mu \nu} &= d_{( \mu} d_{\nu )} \left (Q + \frac{3}{20} R \right )
+ \bigg ( \frac{1}{20} \square R_{\mu \nu} - \frac{1}{15} R_{\mu \alpha} R^{\alpha}_{~ \nu}
+ \frac{1}{30} R_{\mu \alpha \beta \gamma} R_{\nu}^{~ \alpha \beta \gamma} + \nonumber \\
&+ \frac{1}{30}R_{\alpha \beta} R^{\alpha~\beta}_{~\mu~\nu}
\bigg ) + \frac{1}{2} W_{\alpha ( \mu} W^{\alpha}_{~\nu)}, \\
Z_{\mu \nu} &= \mathcal{W}_{\mu \nu} - \frac{1}{2} d_{(\mu} J_{\nu)}, \\
\mathcal{D}_{\mu \nu} &= \mathcal{W}_{\mu \nu} + \frac{1}{2} d_{(\mu} J_{\nu)}.
\end{align}
In the above $T_{(\alpha,\beta)} \equiv \frac{1}{2} \left (T_{\alpha \beta} + T_{\beta \alpha} \right )$
means symmetrization and $O(\nabla^5 g)$ denotes terms with fifth derivative action on the metric tensor 
of the form $\mathcal{R} \nabla \mathcal{R}$. 
The last symbol in definition of $a_4$ is given by
\begin{align}
Z_{(6)} = Z_{(6)}^{M} + Z_{(6)}^S, 
\end{align}
where 
\begin{align}
Z_{(6)}^M &= Q_{(6)} + \frac{5}{2} \left \{ W^{\mu \nu}, \mathcal{R}_{\mu \nu (4)} \right \}
+ \frac{32}{5} \left \{ \mathcal{R}^{\mu \nu \alpha}, \mathcal{R}_{\mu \nu \alpha (2)} \right \}
- \frac{8}{5} \left \{ J^{\mu}, \mathcal{R}_{\mu (4)} \right \}
+ \frac{9}{2} \mathcal{R}^{\mu \nu \alpha \beta} \mathcal{R}_{\mu \nu \alpha \beta} + \nonumber \\
&+ \frac{27}{4} \mathcal{R}^{\mu \nu}_{~~(2)} \mathcal{R}_{\mu \nu (2)}
+ \frac{5}{4} R^{\mu}_{~\nu} \left \{ \mathcal{R}^{\nu \alpha}, \mathcal{R}_{\mu \alpha (2)} \right \}
+ \frac{5}{2} \mathcal{R}_{\mu}^{~ \nu \alpha \beta}\left \{ \mathcal{R}^{\mu \gamma}, \mathcal{R}_{\alpha \beta \nu \gamma} \right \}
+ \nonumber \\
&+ \frac{15}{8} R^{\mu \nu \alpha \beta} \left \{ \mathcal{R}_{\mu \nu}, \mathcal{R}_{\alpha \beta (2)} \right \}
+ \frac{44}{15} d_{\mu} R_{\alpha \nu} \left \{ \mathcal{R}^{\mu \alpha}, J^{\nu} \right \}
+ \frac{22}{5} K^{\mu \nu \alpha \beta \gamma } \left \{ \mathcal{R}_{\mu \gamma}, \mathcal{R}_{\nu \alpha \beta} \right \}
+ \nonumber \\
&+ \frac{22}{5} K_{\mu \nu \alpha (2)} \left \{ \mathcal{R}^{\mu \gamma}, \mathcal{R}^{\nu \alpha}_{~~ \gamma} \right \}
+ \frac{64}{45} R^{\mu}_{~\nu} \mathcal{R}_{\mu \alpha \beta} \mathcal{R}^{\nu \alpha \beta} 
- \frac{16}{15} R^{\mu \nu \alpha \beta} \left \{ d_{\beta} \mathcal{R}_{\mu \nu}, J_{\alpha} \right \}
+ \nonumber \\
&+ \frac{256}{45}R_{\mu ( \alpha | \nu | \beta)} \mathcal{R}^{\mu \alpha}_{~~~ \gamma} \mathcal{R}^{\nu \beta \gamma}
+\frac{32}{45} R^{\mu \nu} J_{\mu} J_{\nu} +
\bigg (
\frac{6}{5} K_{\mu \nu (4)} + \frac{17}{40} R_{\mu \alpha \beta \gamma}R_{\nu}^{~ \alpha \beta \gamma} + \nonumber \\
&+ \frac{17}{60} R_{\mu \alpha} R^{\alpha}_{~ \nu}
\bigg )W^{\mu \sigma}W^{\nu}_{~ \sigma}
+ \bigg (
\frac{24}{5} K_{\mu \nu \alpha \beta (2)} + \frac{17}{40}R_{\mu \gamma}R^{\gamma}_{~ \beta \nu \alpha}
+ \frac{17}{40}R_{\nu \gamma} R^{\gamma}_{~ \alpha \mu \beta} + \nonumber \\
&+ \frac{17}{30} R_{\mu \nu  \sigma \rho}R_{\beta \alpha}^{~~~ \sigma \rho}
+ \frac{17}{60} R_{\mu \sigma \alpha \rho}R^{\sigma~\rho}_{~\nu~\beta}
+ \frac{51}{80} R_{\mu \beta \sigma \rho}R_{\nu \alpha}^{~~~ \sigma \rho}
\bigg )W^{\mu \beta}W^{\nu \alpha}.
\end{align}
The above expression for $Z_{(6)}^M$ is exact in the sense that so far we did not skip any factors
but in concrete calculations many of these terms will produce operators of the order $O(7)$ or $O(\mathcal{R}^3)$
or higher and should be discarded.
Moreover, in the above formula we used
\begin{align}
Q_{(6)} &= g^{\mu_1 \mu_2} g^{\mu_3 \mu_4} g^{\mu_5 \mu_6} d_{(\mu_1} \cdot \cdot \cdot d_{\mu_6)}Q, \\
I^{\alpha \beta}_{~~~\gamma \mu_1 ... \mu_n} &= d_{(\mu_1} \cdot \cdot \cdot d_{\mu_{n-1}} R^{\alpha~\beta}_{~|\gamma| ~\mu_n)}, \\
K^{\alpha \beta}_{~~~ \mu_1 ... \mu_n} &= d_{(\mu_1} \cdot \cdot \cdot d_{\mu_{n-2}} R^{\alpha~\beta}_{~\mu_{n-1} ~\mu_n)}, \\
L^{\alpha}_{~~\mu_1 ... \mu_n} &= d_{(\mu_1} \cdot \cdot \cdot d_{\mu_{n-1}} R^{\alpha}_{ ~\mu_n)}, \\
M_{\mu_1 ... \mu_n} &= d_{(\mu_1} \cdot \cdot \cdot d_{\mu_{n-2}} R_{\mu_{n-1} \mu_n)}, \\
\mathcal{R}^{\mu}_{~ \mu_1 ... \mu_n} &= d_{(\mu_1} \cdot \cdot \cdot d_{\mu_{n-1}} W^{\mu}_{ ~\mu_n)}, \\
I^{\alpha \beta}_{~~~ \gamma \mu (2)} &= g^{\mu_1 \mu_2} I^{\alpha \beta}_{~~~ \gamma \mu \mu_1 \mu_2}, \\
K^{\alpha \beta}_{~~~ \mu \nu (4)} &= g^{\mu_1 \mu_2} g^{\mu_3 \mu_4} K^{\alpha \beta}_{~~~ \mu \nu \mu_1 ...  \mu_4},\\
L^{\alpha}_{~\mu (4)} &= g^{\mu_1 \mu_2} g^{\mu_3 \mu_4} L^{\alpha}_{~\mu \mu_1 ...  \mu_4}, \\
M_{(8)} &= g^{\mu_1 \mu_2} \cdot \cdot \cdot g^{\mu_7 \mu_8} M_{\mu_1 ... \mu_8}, \\
\mathcal{R}^{\mu}_{~ \nu (4)} &= g^{\mu_1 \mu_2} g^{\mu_3 \mu_4} \mathcal{R}^{\mu}_{~ \nu \mu_1 ... \mu_4}.
\end{align}
As far as the $Z_{(6)}^S$ term is concerned, it contains purely gravitational terms of the order $O(\mathcal{R}^3)$ 
or higher, therefore we may put $Z_{(6)}^S = 0$.



\bibliographystyle{JHEP}
\bibliography{cEFT.bib}



\end{document}